\newif\ifconfver
\confvertrue        %declaring conference version u

\ifconfver
\documentclass[10pt,twocolumn,twoside]{IEEEtran}
\else
\documentclass[11pt,draftcls,onecolumn]{IEEEtran}
\fi

\usepackage{makecell}
\usepackage{bbding}
\usepackage{color,graphicx}
\usepackage{float}
\usepackage{multirow,amsmath,epsfig,amsfonts,amsbsy,amssymb,psfig,graphics,psfrag,theorem,calc,url,bm,cite}
\usepackage{stfloats,hyperref}
\usepackage{algorithm,algorithmic}
\usepackage{diagbox} 
\usepackage{hhline}
\usepackage{subfigure}
\usepackage{comment}
\usepackage{tikz}
\usepackage{quantikz}
\usepackage{pdfrender}
\newtheorem{theorem}{Theorem}

\usepackage{amssymb}
\usepackage{pifont}

\usepackage{mathtools}

\makeatletter
\def\multilimits@{\bgroup
	\Let@
	\restore@math@cr
	\default@tag
	\baselineskip\fontdimen10 \scriptfont\tw@
	\advance\baselineskip\fontdimen12 \scriptfont\tw@
	\lineskip\thr@@\fontdimen8 \scriptfont\thr@@
	\lineskiplimit\lineskip
	\vbox\bgroup\ialign\bgroup\hfil$\m@th\scriptstyle{##}$\hfil\crcr}
\def\Sb{_\multilimits@}
\def\endSb{\crcr\egroup\egroup\egroup}
\makeatother

{\begin{list}{}%
		{\setlength{\rightmargin}{0pc}%
			\setlength{\leftmargin}{1pc} }} %
	{\end{list}}

\newlength{\twidth}
\ifconfver
\else
\fi
\definecolor{orange}{RGB}{255,107,0}

\theorembodyfont{\rmfamily}

{\begin{list}{}{
			\settowidth{\labelwidth}{\mbox{\textnormal{#1}}}%
			\setlength{\leftmargin}{\labelwidth+\labelsep}}}%
	{\end{list}}

\newcommand\bE{\ensuremath{{\bm E}}}

\newcommand\bI{\ensuremath{{\bm I}}}

\newcommand\bK{\ensuremath{{\bm K}}}
\newcommand\bL{\ensuremath{{\bm L}}}

\newcommand\bP{\ensuremath{{\bm P}}}

\newcommand\bT{\ensuremath{{\bm T}}}

\newcommand\bV{\ensuremath{{\bm V}}}
\newcommand\bW{\ensuremath{{\bm W}}}
\newcommand\bX{\ensuremath{{\bm X}}}
\newcommand\bY{\ensuremath{{\bm Y}}}

\newcommand\bc{\ensuremath{{\bm c}}}
\newcommand\bd{\ensuremath{{\bm d}}}

\newcommand\bp{\ensuremath{{\bm p}}}
\newcommand\bq{\ensuremath{{\bm q}}}

\newcommand\bs{\ensuremath{{\bm s}}}
\newcommand\bt{\ensuremath{{\bm t}}}

\newcommand\bx{\ensuremath{{\bm x}}}

\definecolor{orange}{RGB}{255,107,0}

\ifconfver
\author{Chia-Hsiang Lin,~\IEEEmembership{Senior Member,~IEEE,}
Jhao-Ting Lin,~\IEEEmembership{Student Member,~IEEE,} \\
Po-Ying Chiu,~\IEEEmembership{Student Member,~IEEE,}
Shih-Ping Chen,
and Charles C. H. Lin}

\title{Quantum-Driven Multihead Inland Waterbody Detection With \\
~~~~~Transformer-Encoded CYGNSS Delay-Doppler Map Data
\thanks{This study was supported by the Emerging Young Scholar Program (namely, the 2030 Cross-Generation Young Scholars Program) of National Science and Technology Council (NSTC), Taiwan, under Grant NSTC 113-2628-E-006-003.
We thank the National Center for Theoretical Sciences (NCTS) and the National Center for High-performance Computing (NCHC) for providing the computing resources.}
\thanks{\textit{(Corresponding author: Shih-Ping Chen.)}}
\thanks{C.-H. Lin is with the Department of Electrical Engineering, and with the Miin Wu School of Computing, National Cheng Kung University, Tainan, Taiwan (R.O.C.) (e-mail: chiahsiang.steven.lin@gmail.com).}
\thanks{J.-T. Lin and P.-Y. Chiu are with the Institute of Computer and Communication Engineering, Department of Electrical Engineering, National Cheng Kung University, Tainan, Taiwan (R.O.C.) (e-mail: q38091534@gs.ncku.edu.tw; f14096148@gs.ncku.edu.tw).}
\thanks{S.-P. Chen and C. C. H. Lin are with the Department of Earth Science, National Cheng Kung University, Tainan, Taiwan (R.O.C.) (e-mail: chensp555@gmail.com; charles@mail.ncku.edu.tw).}
}

\else

\fi

\begin{document}

	\bibliographystyle{IEEEtran}
	\maketitle
	\ifconfver \else \vspace{-0.5cm}\fi

\begin{abstract}
Inland waterbody detection (IWD) is critical for water resources management and agricultural planning.
However, the development of high-fidelity IWD mapping technology remains unresolved.
We aim to propose a practical solution based on the easily accessible data, i.e., the delay-Doppler map (DDM) provided by NASA’s Cyclone Global Navigation Satellite System (CYGNSS), which facilitates effective estimation of physical parameters on the Earth's surface with high temporal resolution and wide spatial coverage.
Specifically, as quantum deep network (QUEEN) has revealed its strong proficiency in addressing classification-like tasks, we encode the DDM using a customized transformer, followed by feeding the transformer-encoded DDM (tDDM) into a highly entangled QUEEN to distinguish whether the tDDM corresponds to a hydrological region.
In recent literature, QUEEN has achieved outstanding performances in numerous challenging remote sensing tasks (e.g., hyperspectral restoration, change detection, and mixed noise removal, etc.), and its high effectiveness stems from the fundamentally different way it adopts to extract features (the so-called quantum unitary-computing features).
The meticulously designed IWD-QUEEN retrieves high-precision river textures, such as those in Amazon River Basin in South America, demonstrating its superiority over traditional classification methods and existing global hydrography maps.
IWD-QUEEN, together with its parallel quantum multihead scheme, works in a near-real-time manner (i.e., millisecond-level computing per DDM).
To broaden accessibility for users of traditional computers, we also provide the non-quantum counterpart of our method, called IWD-Transformer, thereby increasing the impact of this work.
\end{abstract}


\section{Introduction}

Inland waterbody detection (IWD) is a critical surface water monitoring technology which is vital for a range of applications such as agriculture and disaster management.
In particular, accurate detection of surface water distribution is critical for resource assessment and mitigation planning in events such as floods or droughts.
Traditionally, manual surveys have been the most reliable and accurate method for monitoring the status of surface waterbodies.
However, this approach is not feasible for monitoring remote or sparsely populated areas, as the stations require constant maintenance and recalibration, making large-scale near-real-time surveys impractical. 
This limitation is especially evident in regions with dense forests or complex terrain, such as the Amazon rainforest and the jungles of Southeast Asia, where updating stream network databases poses a significant challenge.

Remote sensing technologies, such as optical and radar sensors, have emerged as effective alternatives to ground-based surveys for mapping surface water extent \cite{alsdorf2007measuring}. 
Typically, higher spatial resolution comes with longer revisit cycles and smaller spatial coverage. 
Optical imagery from aircraft or satellites is highly susceptible to cloud cover, precipitation, and vegetation, which can obstruct observations and make it difficult to detect surface waterbodies under adverse weather conditions or dense vegetation. 
To address these challenges and achieve more frequent observations, satellite-based remote sensing technologies have been increasingly employed for monitoring surface water at a global scale. 
One such product is the Global Surface Water Explorer (GSWE), derived from Landsat imagery, which provides global surface water extent data \cite{GSW}.
Additionally, MERIT Hydro is a high-resolution (3-arc second) global hydrography dataset developed from the MERIT DEM \cite{MERIT_DEM} and multiple global waterbody datasets. 
By correcting elevation errors and enhancing the representation of small streams, MERIT Hydro enables more accurate delineation of river networks, thereby supporting a wide range of geoscientific applications \cite{MERIT_Hydro}.

Microwave satellite observations, such as those from Synthetic Aperture Radar (SAR), have also been used to map surface water extent. 
SAR has the advantage of penetrating cloud cover and vegetation, enabling more consistent monitoring of surface water. 
However, interpreting SAR data is complex, as various environmental factors—such as strong winds, rainfall, and vegetation—can affect radar signal reflection and scattering mechanisms, potentially introducing noise into the signal \cite{danklmayer2009assessment}. 
Furthermore, SAR data processing and analysis require significant time, which poses challenges for near-real-time surface water monitoring, especially in applications such as disaster response or seasonal agricultural resource assessment.

Reflectometry is an emerging remote sensing technique that detects surface conditions by analyzing signals reflected from the Earth's surface and received at nadir from Global Navigation Satellite System (GNSS) satellites. 
By examining surface reflectivity, Global Navigation Satellite System Reflectometry (GNSS-R) provides insights to estimate physical parameters at specular points (SPs) based on reflected signals. 
This passive remote sensing approach utilizes existing GNSS signals as signal sources, thereby obviating the need for dedicated transmitters and substantially lowering satellite design costs.
GNSS constellations provide global coverage, which allows GNSS-R to conduct rapid and widespread observations across vast regions. 
Moreover, the long-wavelength L-band signals used in GNSS-R have outstanding penetration abilities, enabling them to pass through cloud cover, rain, snow, and vegetation, thus ensuring reliable, all-weather, and continuous data collection. 
These features make GNSS-R a highly versatile and practical technology, establishing it as a promising tool for Earth sciences and remote sensing. 
The first space-based GNSS-R satellite, the UK Disaster Monitoring Constellation (U.K.-DMC), was launched in 2003 to showcase the viability of space-based GNSS-R \cite{gleason2006remote}. 
TechDemoSat-1 (TDS-1), which was launched in 2014, carried a GNSS-R payload and was retired in 2019 \cite{unwin2016spaceborne}. 
GNSS-R has a broad array of applications in remote sensing, including sea surface wind measurements \cite{hall1988multistatic, martin1993passive, clarizia2009analysis, gleason2005detection, ruf2013cygnss, clarizia2016wind, ruf2018development, cheng2023bagged}, polar sea ice distribution analysis \cite{strandberg2017coastal,zhang2020sea,regmi2022monitoring}, altimetry \cite{wang2022river}, surface water distribution analysis \cite{zhang2022mapping}, and soil moisture content analysis \cite{chew2018soil,chew2020description}.

Launched by NASA in December 2016, the Cyclone Global Navigation Satellite System (CYGNSS) mission comprises a constellation of eight microsatellites employing GNSS-R technology to retrieve sea surface wind speed measurements, with a specific focus on tropical cyclone-prone regions \cite{ruf2018new}.
In addition to providing sea surface wind speed data, the CYGNSS mission has delivered frequent GNSS-R observations of the surface in tropical regions. 
While not originally designed for this purpose, CYGNSS can effectively map surface water extent beneath cloud cover and vegetation \cite{ruf2018new}. 
CYGNSS generates delay-Doppler Maps (DDMs) as part of its Level 1 science data, which can be used to detect surface water distribution and soil moisture. 
The CYGNSS receiver processes reflected signals to create DDMs, representing power values across delay and Doppler frequency ranges. 
However, with a current sampling rate of 2 Hz, each DDM covers approximately 3.5 km of the track, presenting challenges for high-resolution applications. 
The coherence of DDMs is typically assessed by examining the power distribution across the delay and Doppler frequency ranges \cite{al2020algorithm}. 
Through this correlation, several studies \cite{morris2022probabilistic} have explored the relationship between CYGNSS DDM SNR and surface water, further highlighting the platform's potential in IWD. 
Despite this, GNSS-R still faces challenges in spatial resolution compared to traditional remote sensing techniques. 
However, it has demonstrated significant potential for detecting surface water under various weather conditions and in vegetated regions.

\begin{figure*}[t]
    \begin{center}
    \resizebox{1\linewidth}{!}{\hspace{-0cm}\includegraphics{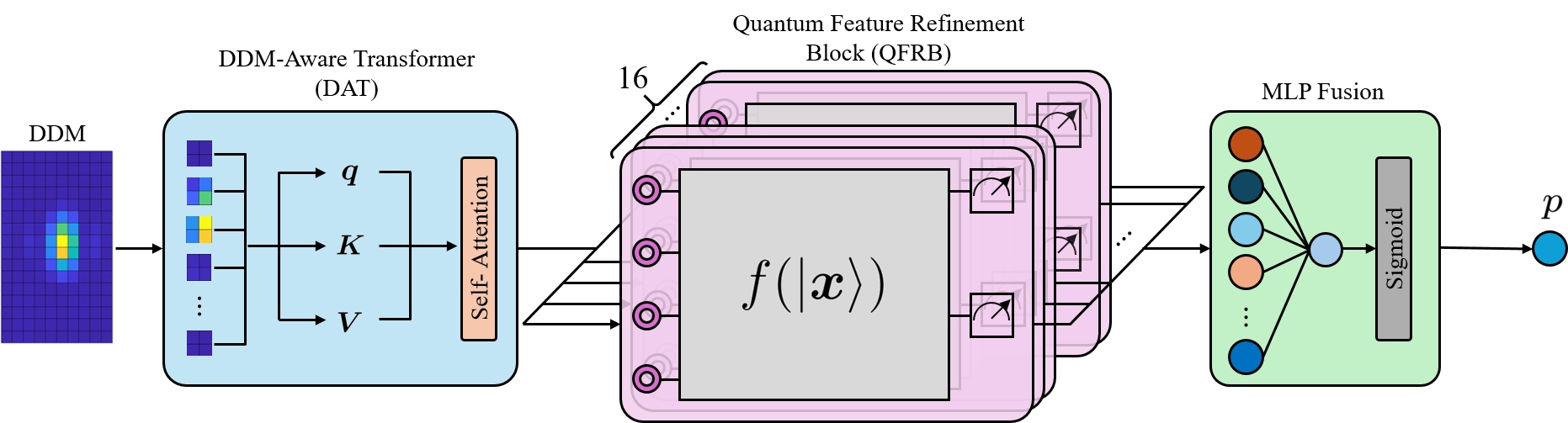}}
    \end{center}
    \caption{Graphical illustration of the proposed IWD-QUEEN for inland waterbody detection. 
    The delay-Doppler map (DDM) is first encoded into concise qubit information by the DDM-aware transformer (DAT), in order to address the issue of limited qubit resources encountered in near-term quantum computers.
    Then, the core quantum signal processing is done by the quantum deep network (QUEEN), called the quantum feature refinement block (QFRB), for extracting quantum features.
    The QUEEN is able to model complex feature interactions in a high-dimensional quantum state space \cite{HyperQUEEN} using a non-shared-weight scheme.
    The detailed architecture will be designed in Figure \ref{fig: detail model}.
    Finally, a simple MLP layer integrates these quantum features to make the final classification decision, returning a sigmoid probability $p$ representing the likelihood of being an inland waterbody region.}
    \label{fig: overall model}
\end{figure*}

To the best of our knowledge, no studies have yet demonstrated a reliable tool or method for detecting surface waterbodies from the GNSS-R DDMs.
Contemporary GNSS-R methodologies frequently encounter limitations in resolving fine-scale hydrological features, such as smaller tributaries, thereby constraining their utility in global-scale hydrological applications.
In this study, we propose an efficient and robust model to monitor target surface waterbodies based on quantum deep network (QUEEN) \cite{HyperQUEEN}, which was originally developed for hyperspectral remote sensing (restoration of NASA's hyperspectral data).
Simply speaking, to achieve effective IWD results, we introduce a highly entangled quantum neural network to analyze trickily encoded DDM information, thereby detecting the waterbody areas.

To be more specific, motivated by the fact that QUEEN has strong proficiency in classification/detection tasks, as evidenced in recent remote sensing literature \cite{QEDNET, QUEEN-G}, we adopt QUEEN to classify DDMs into those that correspond to waterbodies and those that do not.
QUEEN has demonstrated significant potential in solving challenging inference problems across a variety of applications, including hyperspectral tensor completion \cite{HyperQUEEN}, change detection \cite{QUEEN-G}, mangrove mapping \cite{QEDNET}, and drought forecasting \cite{SQUARE-Mamba}.
In particular, recent advances in spectral analysis employ QUEEN to solve the challenging multispectral unmixing, an underdetermined source separation problem, using a quantum light-splitting prism \cite{PRIME}.
Recently, we have also conducted some preliminary investigations of quantum-inspired IWD \cite{HydroQUEEN}.
In this work, we propose IWD-QUEEN, aiming at adopting a more sophisticated transformer technique to elegantly encode the large DDM into compact features, thereby greatly mitigating the issue of limited quantum bits (qubits) in the near-term quantum computers, and then analyze the transformer-encoded features using QUEEN for detecting the waterbody.
The novelties and main contributions of this article are summarized as follows:
\begin{enumerate}
\item 
To the best of our knowledge, this is the first work that proposes a reliable method for high-accuracy detection of surface waterbodies from the GNSS-R DDM data.
In contrast to conventional optical remote sensing techniques, we use DDMs retrieved from the CYGNSS satellite constellation to perform inland waterbody detection (IWD).
By leveraging CYGNSS's superior temporal resolution, our method enables frequent and consistent sampling of waterbody information, thereby enhancing its suitability for near-real-time monitoring of rapid hydrological events (e.g., flood occurrences).
Our method is called IWD-QUEEN.

\item 
For the first time, we explore the potential of quantum deep learning in solving the surface waterbody detection problem.
As this problem can be formulated as a classification problem, QUEEN, having been recognized as a powerful tool for detection/classification missions, does achieve remarkable IWD results, especially when identifying those fine details of river textures.

\item 
As directly processing large DDMs is not friendly to near-term quantum computers (whose qubit resources are limited), we, inspired by the deep compression module in \cite{HyperQUEEN}, propose to analyze the encoded/compressed DDMs.
Accordingly, the designed DDM-aware transformer (DAT) implements a learnable fusion between the classification tokens and the features extracted from the key region of the DDM. 
This fusion strategy facilitates the adaptive integration of global contextual representations with regionally focused physical cues, improving the model’s sensitivity to waterbody-specific patterns and hence facilitating the subsequent classification.

\item 
The designed QUEEN architecture is novel, with the design philosophy applicable to other related fields.
Specifically, in addition to the mathematically guaranteed quantum expressibility (cf. Theorem \ref{theorem: fully expressibility}) of the IWD-QUEEN, we further enhance its representational capacity by integrating strongly entangled quantum circuits, which introduce high levels of entanglement among qubits.
In addition, the non-shared-weight mechanism allows us to independently analyze different quantum features (i.e., quantum multihead mechanism), enabling independent feature transformation and enhancement.
This mechanism helps better discriminate the surface waterbody and land areas, and yields accurate IWD outcomes (cf. \ref{subsec: qq_analysis}) in a near-real-time manner (i.e., millisecond-level computing per DDM).

\item 
To facilitate accessibility to general users and to increase the impact of this work, we also propose a non-quantum counterpart of our method, called IWD-Transformer, which can be accessed and implemented on classical computers.
\end{enumerate}

The remainder of this article is organized as follows.
In Section \ref{sec: method}, we design the proposed IWD-QUEEN and IWD-Transformer. 
Section \ref{sec: experimental results} presents and discusses extensive experimental results conducted over the Amazon River Basin in South America, including quantitative evaluations and ablation studies.
In addition, several case studies, covering varying seasonal conditions, human activities, and the 2025 seismic event, are performed over the Amazon River Basin and the Mekong River Basin in Southeast Asia, illustrating the practical applicability and generalization capability of IWD-QUEEN under real-world conditions. 
Finally, Section \ref{sec: conclusion} concludes this work.

Some standard notations used in this article are collectively introduced hereinafter.
$\mathbb{R}^{n}$ denotes the set of real-valued $n$-dimensional vectors, and $\mathbb{R}^{m\times n}$ denotes the set of real-valued $m$-by-$n$ matrices.
$\text{DIAG}(\bY_1, \dots,\bY_N)$ denotes the block-diagonal matrix with $\bY_n$ being the $n$th diagonal block \cite{CVXbookCLL2016}.
$\bI_m$ represents the $m$-by-$m$ identity matrix.
Some more advanced quantum-related notations will be introduced later.

\begin{figure*}[t]
    \begin{center}
    \resizebox{1\linewidth}{!}{\hspace{-0cm}\includegraphics{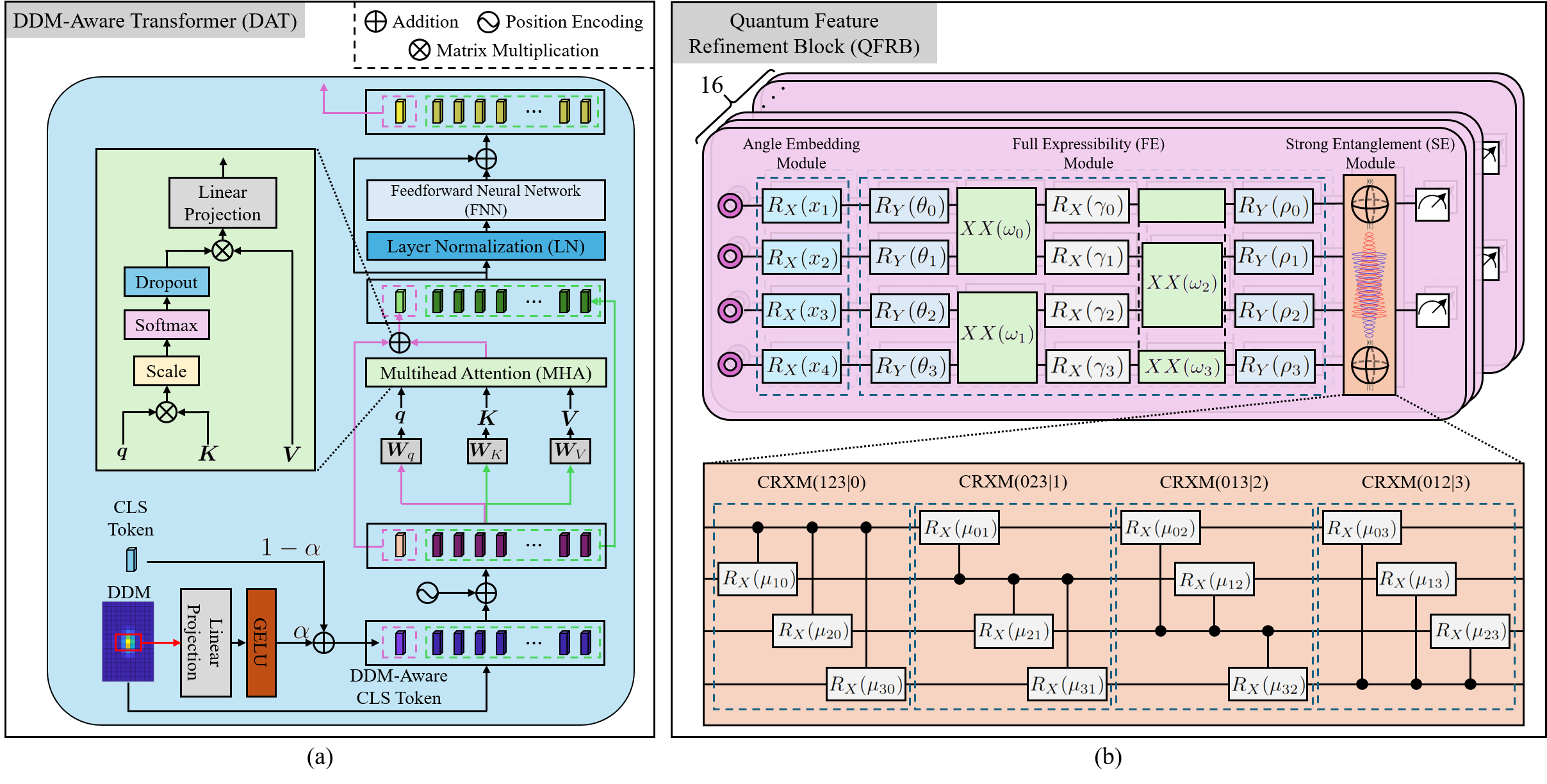}}
    \end{center}
    \caption{Detailed architectures of (a) the proposed DDM-aware transformer (DAT) and (b) the quantum feature refinement block (QFRB).
    In (a), the DAT is designed to fuse the global contextual information from the CLS token with the physical attributes extracted from the DDM.
    The resulting DDM-aware CLS token is subsequently passed to a transformer-based encoder to capture spatial dependencies.
    In (b), the QFRB performs feature enhancement from the $16$ deep features $\bx_1,\bx_2,\cdots,\bx_{16}$.
    We remark that $x_i$’s are not learning parameters; instead, they serve as embedded data inputs, meaning that $x_i$ is encoded as a rotation angle in a quantum $R_X$ gate (i.e., quantum angle embedding). 
    Then, the quantum features are fed into the trainable FE module, which operates with distinct parameter sets for each quantum circuit and can provably realize any valid quantum operators (cf. Theorem \ref{theorem: fully expressibility}).
    Once the quantum signal processing is done with the FE module, each feature vector is then processed through the SE module, which adopts a strongly entangled quantum circuit structure (four controlled rotation X modules (CRXMs)) to enhance individual feature representations.
    }
    \label{fig: detail model}
\end{figure*}

\section{Proposed IWD Methods} \label{sec: method}

Motivated by the limitations mentioned above, we propose IWD-QUEEN, a novel hybrid classical-quantum framework designed to fully harness the entanglement capabilities of QUEEN in a cascaded manner, thereby extending its applicability to hydrological remote sensing.
To facilitate the readability, we outline its technical design hereinafter.
First, we design a DDM-aware transformer (DAT), which incorporates a learnable convex combination between the traditional classification (CLS) token and physical attributes extracted from the DDM, particularly emphasizing its central region.
This design is informed by the observation that, due to coherent scattering effects, the DDM signatures associated with inland waterbodies are often highly concentrated around the central region, making this area particularly informative for the subsequent classification.
The transformer-encoded DDM (tDDM) are then embedded as the angles of very few quantum gates (echoing the limited qubit resources in near-term quantum computers \cite{IBM}), and that quantum angle information is then processed using a customized quantum feature refinement block (QFRB) wherein the core quantum signal processing is done.
By assigning distinct parameter sets to each feature vector, QFRB enables tailored feature transformation and individualized representation learning.
The designed QFRB is theoretically proven to achieve quantum full expressibility (FE) to efficiently represent the features under a lightweight quantum architecture.
We also incorporate strongly entangled quantum circuits in QFRB by using controlled rotation X gates to capture intricate feature correlations.
Finally, the refined quantum features are passed through a simple multi-layer perceptron (MLP) layer, which fuses the quantum information features to estimate the likelihood of the presence of inland waterbodies.
The above design philosophy of IWD-QUEEN is graphically illustrated in Figure \ref{fig: overall model}.

Given the strong performance of IWD-QUEEN, some users might be interested in conducting waterbody detection on classical computers.
For this category of users, we develop a non-quantum counterpart of IWD-QUEEN, termed IWD-Transformer, thereby increasing the impact of our work.
In IWD-Transformer, the QFRB (the core quantum network of IWD-QUEEN) is replaced with an equally lightweight convolutional neural network (CNN) architecture.
As it turns out, albeit with some performance compromises, IWD-Transformer still has outstanding detection performance on real CYGNSS data (cf. Section \ref{subsec: qq_analysis}).

The remaining parts of Section \ref{sec: method} are organized as follows to clearly provide a comprehensive description of the IWD-QUEEN (along with its core design philosophy) and IWD-Transformer.
To this end, the overall architecture of IWD-QUEEN can be mathematically formulated to return the probability $p$ of being a waterbody region, i.e.,
\begin{align} 
p = \text{MLP}\left( \text{QFRB}\left( \text{DAT}\left(\bX \right) \right) \right) \in [0,1], 
\end{align}
where $\bX$ denotes the input DDM of the target region, whose foundational principles will be introduced in Section \ref{subsec: DDM}.
The operator $\text{DAT}(\cdot)$ refers to the DDM-aware transformer function, which adaptively fuses global contextual features with localized physical characteristics extracted from $\bX$; further architectural details are provided in Section \ref{subsec: DAT}. 
The tDDM from DAT is then processed by $\text{QFRB}(\cdot)$, a QUEEN featuring a non-shared-weight mechanism and strong entanglement, designed to enhance inter-feature correlations and refine the global representation; this component is described in Section \ref{subsec: QFRB}. 
Finally, $\text{MLP}(\cdot)$ denotes a conventional multi-layer perceptron layer that projects the learned representation into the output prediction result $p \in [0,1]$, which corresponds to the probability of waterbody presence. 
The overall workflow of the IWD-QUEEN is graphically summarized in Figure \ref{fig: overall model}.
The detailed structure of DAT is illustrated in Figure \ref{fig: detail model}(a), while QFRB is depicted in Figure \ref{fig: detail model}(b).
Besides, Section \ref{subsec: loss} discusses the loss function used for addressing the serious label imbalance issue encountered in the IWD problem \cite{LIIP}.
Finally, Section \ref{subsec: nonquantum} introduces the classical counterpart of the IWD-QUEEN (i.e., the aforementioned IWD-Transformer) to facilitate broader accessibility of our proposed methods.

\subsection{Delay-Doppler Map (DDM)} \label{subsec: DDM}

\begin{figure}[!t]
    \centering
    \resizebox{1\linewidth}{!}{%
        \includegraphics{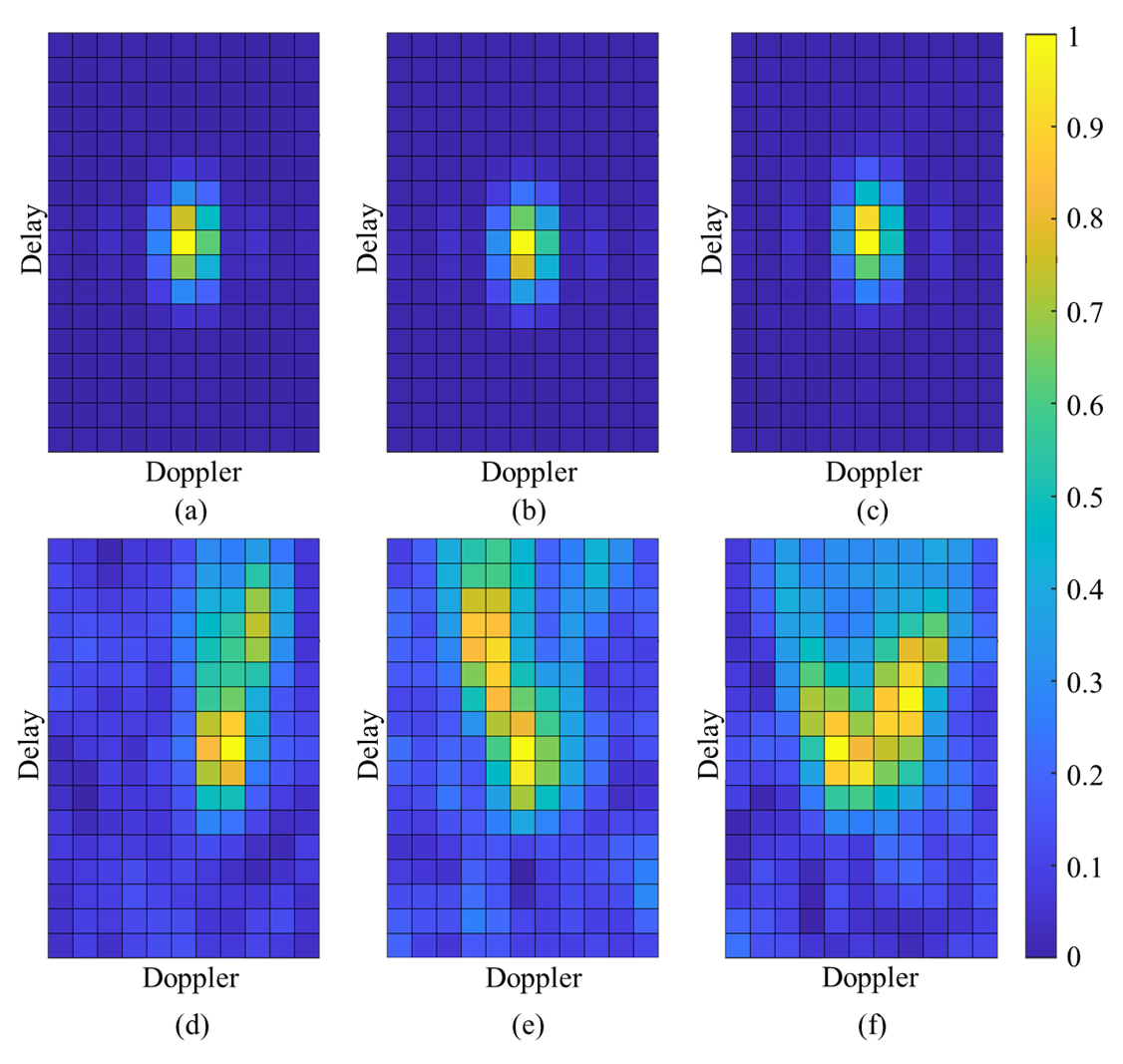}
    }
    \caption{Delay-Doppler maps (DDMs) of inland waterbodies and land surfaces.
    (a)-(c) illustrate DDMs of inland waterbodies, which are characterized by concentrated signal energy.
    In contrast, (d)–(f) present DDMs of land surfaces, exhibiting more dispersed and irregular energy distributions.
    }
    \label{fig: six ddm}
\end{figure}

The DDM Instrument (DDMI) is the sole payload aboard each of the eight individual satellites comprising the CYGNSS constellation \cite{gleason2016calibration}.
Its main function is to receive Global Positioning System (GPS) signals reflected from the Earth’s surface and convert them into DDMs, which are used to characterize the scattering properties of the surface.
A DDM is a two-dimensional matrix that represents the distribution of signal energy across different time delays and Doppler frequencies, offering insight into the surface roughness.
Smooth water surfaces tend to produce coherent, mirror-like reflections, resulting in sharp, localized peaks in the DDM, as illustrated in Figures \ref{fig: six ddm}(a)-\ref{fig: six ddm}(c).
In contrast, rough surfaces such as vegetation, soil, or built environments typically lead to incoherent scattering patterns (cf. Figures \ref{fig: six ddm}(d)–\ref{fig: six ddm}(f)) that appear to have more dispersed energy in the DDM. 

In this article, we utilize the version 3.2 geo-located DDMs calibrated into power received (Watts) in CYGNSS Level 1 science data record dataset \cite{CYGNSS_Handbook}.
Each DDM consists of 17 delay bins (sampled at 0.25 $\mu$s) and 11 Doppler bins (sampled at 500 Hz), reported at a 2 Hz interval \cite{9328203}.
This leads to the $17\times 11$ matrix representation of each DDM, as illustrated in Figure \ref{fig: six ddm}.
With each satellite capable of simultaneously tracking four SPs, CYGNSS can generate up to 64 DDMs per second. 
This high temporal resolution and frequent revisit time (several hours) make CYGNSS ideal for capturing short-term changes in the areas of inland waterbodies.

\subsection{DDM-Aware Transformer (DAT)} \label{subsec: DAT}
The DDM produced by CYGNSS can be represented as a two-dimensional matrix, where one axis corresponds to signal time delay and the other to Doppler frequency. 
Each matrix element, or bin, indicates the power intensity associated with a specific delay-Doppler pair.
This matrix structure resembles an image with spatially meaningful patterns so that we can use a very lightweight transformer for the feature extraction process, which is partly enabled by the small scale and the single-channel nature of DDMs, and partly attributed to the relative simplicity of the binary classification task.
By considering these factors, the computational burden of encoding the DDMs can be significantly reduced, allowing a concise modeling of DDM's global dependencies with low architectural complexity (the so-called lightweight AI).

Given a DDM $\bX \in \mathbb{R}^{H \times W}$, where $H := 17$ and $W := 11$ represent the height and width, respectively.
The matrix $\bX$ is initially partitioned into non-overlapping patches to facilitate the subsequent transformer-driven encoding procedure.
These patches are subsequently flattened and organized into  $\bP = [ \bp_1, \bp_2,\cdots, \bp_N ]^T \in \mathbb{R}^{N \times P^2}$, where $\bp_i$ denotes the $i$th patch vector, $N = HW / P^2$ is the total number of patches, and the patch size is $P\times P$.
As $H$ and $W$ are odd, the last row and last column of the DDM are omitted (typically not containing significant information), and accordingly we set $P \coloneqq 2$ in this article.
The flattened patches in $\bP$ are then linearly projected into a feature vector of size 64 through a projection matrix $\bL \in \mathbb{R}^{P^2 \times 64}$, yielding a sequence of patch tokens. 
In addition, a CLS token $\bc \in \mathbb{R}^{64}$ is prepended \cite{ViT, TGFA-AD}, and positional information is incorporated via a position embedding matrix $\bE \in \mathbb{R}^{(N+1) \times 64}$. 
Therefore, the initial input to the transformer encoder can be defined as
\begin{align} 
\bT = \left( \bc^T \| \bP \bL \right) + \bE \in \mathbb{R}^{(N+1) \times 64}, 
\end{align}
where the notation ``$\|$'' denotes the concatenation operation along the row dimension.

To improve the model's effectiveness in classifying inland waterbodies, we propose a novel DDM-aware transformer (DAT).
Rather than directly forwarding the original CLS token $\bc$ to the subsequent layers, we introduce the DDM-aware CLS token $\bc_{\text{DDM}} \in \mathbb{R}^{64}$, which incorporates spatial priors from the central region of $\bX$.
This region, represented as $\bX_{\text{central}} \in \mathbb{R}^{h \times w}$ (typically, $(h,w) \coloneqq (3,5)$ \cite{9328203}), is empirically selected based on the observation that it often contains the most discriminative signal characteristics.
In particular, DDMs corresponding to inland waterbodies often exhibit concentrated signal energy in the central region due to coherent scattering, whereas DDMs associated with land surfaces typically display more dispersed and irregular energy distributions.
To extract a representative feature $\bd \in \mathbb{R}^{64}$ from the central region, a linear projection (LP) is applied to the flattened $\bX_{\text{central}}$ (i.e., $\bx_{\text{central}}$), followed by a Gaussian error linear unit (GELU) activation function, yielding
\begin{align} \label{d}
\bd = \text{GELU}\left( \text{LP}\left( \bx_{\text{central}} \right) \right).
\end{align}
The final DDM-aware CLS token $\bc_{\text{DDM}}$ is formed using a learnable convex combination of the original CLS token $ \bc$ and the extracted central feature $\bd$, as defined below
\begin{align} 
\bc_{\text{DDM}} = \left( 1-\alpha \right) \bc+ \alpha \bd,
\end{align}
where $\alpha\in[0,1]$ is a trainable scalar parameter defining a specific convex combination \cite{CVXbookCLL2016}.
This design encourages the CLS token to better capture the distinctive feature of inland waterbodies, which typically exhibit high-intensity responses in the center of the DDM.
The DDM-informed token sequence $\bT_{d}$ is formulated as:
\begin{align} 
\bT_d = \left( \bc_{\text{DDM}}^T \| \bP \bL \right) + \bE \in \mathbb{R}^{(N+1) \times 64}.
\end{align}
Subsequently, each token in $\bT_{d}$ undergoes layer normalization (LN) before being processed by the multihead attention (MHA) mechanism.

The MHA consists of several self-attention (SA) heads that operate in parallel, allowing the model to simultaneously attend to information from different representation subspaces (i.e., different heads) across various spatial positions \cite{TGFA-AD}.
The DDM-aware CLS token with positional encoding is denoted as $\bt_{\text{CLS}} \in \mathbb{R}^{1 \times 64}$ and is obtained from the first row of $\bT_d$. 
The remaining rows, corresponding to the patch tokens, constitute the local representation matrix $\bT_{\text{local}} \in \mathbb{R}^{N \times 64}$.
To facilitate attention computation and further reduce the network complexity, only $\bt_{\text{CLS}}$ is projected into a query vector $\bq$, while $\bT_{\text{local}}$ is linearly mapped to key matrix $\bK$ and value matrix $\bV$, using trainable projection matrices $\bW_q, \bW_K, \bW_V \in \mathbb{R}^{64 \times 64}$, respectively, and they can be formulated as
\begin{align}
\bq = \bt_{\text{CLS}}\bW_q ,~\bK = \bT_{\text{local}} \bW_K,~\bV = \bT_{\text{local}} \bW_V. 
\end{align}
The attention output from $\text{MHA}(\cdot)$ captures the global contextual relevance of each patch token with respect to $\bq$ and is formulated as
\begin{align} 
\text{MHA} \left( \bT_{d} \right) = \text{Softmax}\left( \frac{\bq \bK^T}{\sqrt{D}} \right) \bV, 
\end{align} 
where $D$ is a scaling factor to ensure stable gradient flow during the training phase.

The above mechanisms are graphically illustrated in Figure \ref{fig: detail model}(a), where the overall procedure of the DAT function is defined as
\begin{align} \label{DAT}
\bT_{d}'&= \left( \text{MHA} \left(\bT_{d} \right)+ \bt_{\text{CLS}} \right) \| \bT_{\text{local}},\notag
\\
\text{DAT}(\bX) &= \bT_{d}'+ \text{FNN} \left( \text{LN} \left( \bT_{d}' \right) \right),
\end{align}
where FNN denotes the feedforward neural network, which can be defined as
\begin{align} \label{FNN}
\text{FNN}(\cdot) = \text{DO}\left( \text{LP}\left( \text{DO}\left(  \text{GELU}\left( \text{LP}\left(\left( \cdot \right) \right) \right) \right) \right) \right),
\end{align}
where $\text{DO}(\cdot)$ means the dropout operator.
Finally, we extract the first row of the output in \eqref{DAT}, serving as the DAT output $\bt_{\text{DDM}} \in \mathbb{R}^{64}$ for the following quantum signal processing, where $\bt_{\text{DDM}}$ has been sufficiently compact for mitigating the issue of limited qubit resources encountered in the near-term quantum computers.

\subsection{Design of IWD-QUEEN} \label{subsec: QFRB}
Just as human observers tend to intuitively evaluate DDMs based on key features such as signal concentration, spatial relationships along the delay and Doppler axes, or directional gradients, various aspects for assessing the DDMs may reveal distinct and valuable patterns for IWD.
Inspired by this perspective, $\bt_{\text{DDM}}$ is reshaped into 16 separate deep feature vectors $\bm{x}_1, \bm{x}_2, \dots, \bm{x}_{16} \in \mathbb{R}^4$.
Each vector is then fed into an individual quantum circuit (i.e., with non-shared weights) working as an independent quantum head, which enables tailored feature transformation and individualized representation learning.
To be more specific, we adopt quantum angle embedding \cite{HyperQUEEN}, wherein each vector $\bm{x}_i$ is encoded into a 4-qubit quantum state by the quantum rotation gate $R_{X}$ (defined in Table \ref{tab: common_qu_gate}).
To further enhance the extract features, a customized quantum feature refinement block (QFRB) is proposed, wherein the core quantum signal processing is conducted.
As schematically shown in Figure \ref{fig: detail model}(b), the overall procedure of the QFRB, denoted by $\text{QFRB}(\cdot)$, is defined as
\begin{align} \label{QFRB}
\text{QFRB}(\cdot) = \text{SE}\left( \text{FE}\left( \cdot \right) \right),
\end{align}
where FE refers to the ``full expressibility'' module, and SE represents the ``strong entanglement'' module, as will be designed and described next.

The FE module (cf. the upper region in Figure \ref{fig: detail model}(b)) adopts the quantum processing strategy proposed in the hyperspectral QUEEN (HyperQUEEN) \cite[Theorem 2]{HyperQUEEN}, which has demonstrated effectiveness in a variety of remote sensing tasks, including change detection \cite{QUEEN-G}, mangrove mapping \cite{QEDNET}, drought forecasting \cite{SQUARE-Mamba}, and multispectral unmixing (mathematically equivalent to the underdetermined problem), and has achieved outstanding performance.
This module consists of a sequential arrangement of quantum gates, i.e., rotation Y, Ising XX, rotation X, Ising XX, and rotation Y, as depicted in Figure \ref{fig: detail model}(b).
This sequential construction mathematically ensures that the FE module can efficiently represent the features under a lightweight quantum architecture, because it uses only 5 layers while being able to fully express any valid quantum operator $U$, as guaranteed in the following theorem.
\begin{theorem}
\textit{The trainable quantum neurons deployed in the FE module (cf. Figure \ref{fig: detail model}(b)) can jointly express any valid quantum unitary operator \( U \), with some real-valued trainable network parameters \( \{\theta_k, \omega_k, \gamma_k, \rho_k\} \), where $\theta_k$ (resp., $\rho_k$) is the angle of the first quantum rotation Y gate (resp., the last quantum rotation Y gate), $\omega_k$ is the angle of the Ising quantum gate, and $\gamma_k$ is the angle of the quantum rotation X gate.}
\hfill $\square$
\label{theorem: fully expressibility}
\end{theorem}
The proof of Theorem \ref{theorem: fully expressibility} closely aligns with the proof procedure outlined in \cite{HyperQUEEN}, and is omitted here for conciseness.

As for the SE module, illustrated in the lower part of Figure \ref{fig: detail model}(b), it is motivated by the fact that strong quantum entanglement has been found effective in classification/discrimination-like missions \cite{HyperKING}.
For IWD, which can be formulated as a classification problem, we upgrade the entanglement status after ensuring the FE property (i.e., Theorem \ref{theorem: fully expressibility}).

In our design, the SE module incorporates four controlled rotation X modules (CRXMs) to capture intricate feature correlations among the entangled qubits.
To be rigorous, each qubit in CRXM is entangled with all others through the controlled rotation X (CRX) gates, denoted as $\text{CRXM}({ijk}|{l})$, where $l$ denotes the control qubit and $i, j, k$ are the target qubits. 
For example, the notation $\text{CRXM}({123}|{0})$ indicates that qubit 0 serves as the control for CRX gates applied to qubits 1, 2, and 3; this is graphically illustrated in the lower part of Figure \ref{fig: detail model}(b).
This configuration enables full pairwise entanglement across the qubits, allowing the quantum circuit to encode essential quantum correlations that are critical for the downstream decision-making procedure.

The quantum gates adopted in both FE and SE modules, along with their corresponding symbols and mathematical definitions, are summarized in Table \ref{tab: common_qu_gate}.
We remark that, unlike their classical counterparts, which are predominantly irreversible, the quantum gates (i.e., quantum neurons) are inherently reversible due to their unitary nature, enabling end-to-end training via optimizers such as Adam \cite{kingma2014adam}.
Given that current quantum devices operate within the noisy intermediate-scale quantum (NISQ) regime \cite{HyperQUEEN}, one practical challenge involves managing the quite limited number of qubits available. 
In our architecture, 16 independently parameterized quantum heads are employed in the proposed IWD-QUEEN, each using 4 qubits, leading to a total of only 64 qubits.
This scale remains within the capabilities of modern quantum hardware, such as IBM's 433-qubit ``Osprey" introduced in 2022 \cite{HyperQUEEN} or the recently announced ``Condor" processor, which offers up to 1121 qubits \cite{castelvecchi2023ibm}.
This ensures that the proposed quantum model remains both computationally feasible and implementable with currently available technologies while maintaining the advantages of quantum parallelism and entanglement.

\begin{table}[t]
\scriptsize
\setlength{\tabcolsep}{4pt} % 
\caption{Quantum neurons (or quantum gates) used in the proposed IWD-QUEEN, where $\delta \triangleq \cos(\theta/2)$ and $\gamma \triangleq \sin(\theta/2)$.}\label{tab: common_qu_gate}
\vspace{-0.3cm}
\begin{center}
\begin{tabular}{|c c c|} 
 \hline
 \rule{0pt}{2ex}
 Quantum Neuron & Symbol & Unitary Operator 
 \rule{0pt}{2ex}
 \\
 \hline
 \rule{0pt}{4ex}
 Rotation X
 &
 \begin{tikzcd}
    \qw & \gate{R_{X}(\theta)} & \qw
 \end{tikzcd}
 &
 $\begin{pmatrix}
    \delta & -i \gamma \\
    -i \gamma & \delta
\end{pmatrix}$
 \rule{0pt}{4ex}
 \\ 
 \hline
 \rule{0pt}{4ex}
 Rotation Y
 &
 \begin{tikzcd}
    \qw & \gate{R_{Y}(\theta)} & \qw 
 \end{tikzcd}
 & 
 $\begin{pmatrix}
    \delta & - \gamma \\
    \gamma & \delta
\end{pmatrix}$
 \rule{0pt}{4ex}
%\rule[-5ex]{0pt}{4ex}
 \\
 \hline
 \rule{0pt}{4ex}
 \rule{0pt}{6.5ex}
 Ising XX
&
\begin{quantikz}
    \qw & \gate{XX(\theta)} & \qw
\end{quantikz}
&
$\begin{pmatrix}
    \delta & 0 & 0 & -i\gamma\\
    0 & \delta & -i\gamma & 0\\
    0 & -i\gamma & \delta & 0\\
    -i\gamma & 0 & 0 & \delta
\end{pmatrix}$
\\
\hline
 %\rule{0pt}{11ex}
Controlled $R_X$
 &
\begin{quantikz}
    & \ctrl{1} & \qw \\
    & \gate{R_X(\theta)} & \qw
\end{quantikz}
 &
 $\textrm{DIAG}(\bm{I}_2,R_{X}(\theta))$
 \rule[-5ex]{0pt}{4ex}
 \\
 \hline
 Pauli-Z &
\begin{tikzcd}
\meter{} 
\end{tikzcd}
&
$\begin{pmatrix}
    1 & 0 \\
    0 & -1 \\
\end{pmatrix}$
%\rule[-5ex]{0pt}{4ex}
\\
\hline
\end{tabular}
\vspace{-0.7cm}
\end{center}
\end{table}

\begin{center}
\scriptsize
\begin{table}[t]
    \caption{Implementation details of QFRB and MLP in IWD-QUEEN.}
    \centering
    \label{tab: quantum model table}
    
\begin{tabular}{c||c|c|c}
    \hline
    \hline
    \rule{0pt}{2.3ex}
     & Layer & Configuration & Output Size
    \rule{0pt}{2ex}
    \\
    \hline
    \hline
    \rule{0pt}{2ex}
     & Input & - & 16$\times$4 
    \rule{0pt}{2ex} \\
    \hline
    \multirow{11}{*}{\makecell{FE Module\\(\textbf{Theorem \ref{theorem: fully expressibility}})}} & Angle Embedding  & $R_{X}$(16, 4) & 16$\times$4
    \rule{0pt}{2.3ex}\\
    \cline{2-4} & Unitary Gate 1 & $R_{Y}$(16, 4) & 16$\times$4
    \rule{0pt}{2.3ex}\\
    \cline{2-4} & Reshape & - & 32$\times$2
    \rule{0pt}{2.3ex}\\
    \cline{2-4} & Unitary Gate 2 & $XX$(16, 2) & 32$\times$2
    \rule{0pt}{2.3ex}\\
    \cline{2-4} & Reshape & - & 16$\times$4
    \rule{0pt}{2.3ex}\\
    \cline{2-4} & Unitary Gate 3 & $R_{X}$(16, 4) & 16$\times$4
    \rule{0pt}{2.3ex}\\
    \cline{2-4} & Reshape & - & 32$\times$2
    \rule{0pt}{2.3ex}\\
    \cline{2-4} & Unitary Gate 4 & $XX$(16, 2) & 32$\times$2
    \rule{0pt}{2.3ex}\\
    \cline{2-4} & Reshape & - & 16$\times$4
    \rule{0pt}{2.3ex}\\
    \cline{2-4} & Unitary Gate 5 & $R_{Y}$(16, 4) & 16$\times$4
    \rule{0pt}{2.3ex}\\
    \hline
    \multirow{4}{*}{\makecell{SE Module}} & \multirow{4}{*}{\makecell{Unitary\\Gates}} & CRXM(123$|$0) & 16$\times$4
    \rule{0pt}{2.3ex} \\
    \cline{3-4}
    & & CRXM(023$|$1) & 16$\times$4 \rule{0pt}{2.3ex} \\
    \cline{3-4}
    & & CRXM(013$|$2) & 16$\times$4 \rule{0pt}{2.3ex} \\
    \cline{3-4}
    & & CRXM(012$|$3) & 16$\times$4 \rule{0pt}{2.3ex} \\
    \hline
    \multirow{2}{*}{Measurement} & \rule{0pt}{4ex} \makecell{Pauli-Z\\Measurement} \rule{0pt}{4ex}& $Z$(16, 2) & 16$\times$2 \rule{0pt}{2.3ex} \\
    \cline{2-4}
    & Reshape & - & 32$\times$1 \rule{0pt}{2.3ex} \\
    
    \hline
    \multirow{5}{*}{\makecell{Fusion}} & \multirow{5}{*}{MLP} & LP(32, 16) & 16$\times$1 \rule{0pt}{2.3ex} \\
    \cline{3-4}
    & & GELU & 16$\times$1 \rule{0pt}{2.3ex}\\
    \cline{3-4}
    & & DO(0.1) & 16$\times$1 \rule{0pt}{2.3ex}\\
    \cline{3-4}
    & & LP(16, 1) & 1$\times$1 \rule{0pt}{2.3ex}\\
    \cline{3-4}
    & & Sigmoid & 1$\times$1 \rule{0pt}{2.3ex}\\
    \hline
\end{tabular}
\end{table}
%\vspace{-0.5cm}
\end{center}

Finally, to mimic the max pooling in the traditional deep learning, only half of the 64 qubits are measured, leading to the refined quantum features $\bs \in \mathbb{R}^{32}$.
Specifically, the refined quantum features $\bs$ are derived by performing quantum measurements on two specific qubits (i.e., the first and third qubits) for each of the 16 independent heads, as graphically illustrated in the upper part of Figure \ref{fig: detail model}(b).
Overall, the above scheme can be written as
\begin{equation}
\bs = [\bs_1^T, \bs_2^T, \dots, \bs_{16}^T]^T,
\end{equation}
where each component $\bs_i \in \mathbb{R}^2$ represents the measured result extracted from the first/third qubits of the $i$th quantum circuit (i.e., the $i$th quantum head).

Subsequently, to estimate the probability $p$ of inland waterbody presence, a simple MLP fusion layer is employed to fuse the quantum information features for final classification, mathematically formulated as 
\begin{align} \label{MLP}
\text{MLP}(\cdot) = \text{Sigmoid}\left( \text{LP}\left( \text{DO}\left( \text{GELU}\left( \text{LP}\left( \cdot \right) \right) \right) \right) \right),
\end{align}
where ``$\text{Sigmoid}(\cdot)$'' denotes the sigmoid function.
A binary decision is then derived by applying a threshold $t := 0.5$ to the predicted probability $p$, thereby effectively distinguishing between inland waterbody and land surface classes. 

To further support reproducibility and clarity, the detailed structures of QFRB and MLP fusion (including their configuration and output size of each layer) are provided in Table \ref{tab: quantum model table}, where ``$R_Y(g, n)$," ``$R_X(g, n)$," ``$XX(g, n)$," and ``$Z(g, n)$" are the unitary quantum gates, in which $g$ and $n$ respectively denote the number of parallel groups and the number of qubits in each group.
``$\text{LP}(a, b)$" denotes linear projection operation with $a$ and $b$ being the input dimension and output dimension, respectively, and ``$\text{DO}(c)$" is the dropout operation with $c$ being the dropout rate.
Therefore, we have completed the design of IWD-QUEEN algorithm.

\subsection{Loss Function} \label{subsec: loss} 
Since the problem is a binary classification task, the commonly used binary cross-entropy (BCE) loss \cite{de2005tutorial} is naturally adopted.
BCE loss measures the dissimilarity between the predicted probability \( p \in [0, 1] \) and the ground-truth label \( p_{\text{GT}} \in \{0, 1\} \), which can be defined as 
\begin{equation}
\mathcal{L}_{\text{BCE}} = -\frac{1}{B} \sum_{i=1}^{B} \left[
p_{\text{GT},i} \log(p_i) + (1 - p_{\text{GT},i}) \log(1 - p_i)
\right]
\end{equation}
where $B$ denotes the batch size.
Furthermore, considering the serious label imbalance issue commonly encountered in the IWD problem \cite{LIIP}, where samples of land surfaces significantly exceed those of inland waterbodies, we incorporate Kappa loss \cite{kappa_loss} as an additional training loss for mitigating the adverse effects during model optimization.
Specifically, Cohen’s Kappa coefficient ($\kappa$) \cite{kappa} is employed as a loss function $\mathcal{L}_{\kappa}$ due to its capability for assessing agreement in binary classification tasks.
Overall, the loss function for training IWD-QUEEN and IWD-Transformer is defined as
\begin{equation}
\mathcal{L}_{\text{total}} = \mathcal{L}_{\text{BCE}} + \mathcal{L}_{\kappa},
\end{equation}
whose implementation will be discussed next.

Note that Cohen’s Kappa is not directly implementable.
Specifically, in the naive definition of Cohen’s Kappa, the use of non-differentiable discrete hard labels will hinder the effective gradient descent.
To address this limitation, Kappa loss $\mathcal{L}_\kappa$ employs probability values as a soft approximation of $\kappa$, enabling gradient-based optimization and facilitating the training process.
To be rigorous, the formulation of $\mathcal{L}_\kappa$ is explicitly defined as \cite{kappa_loss}
\begin{equation}
\mathcal{L}_\kappa = 1 - 
\frac{
2 \sum_{i=1}^{B} p_i p_{\text{GT},i} - \left( \sum_{i=1}^{B} p_i \right) \left( \sum_{i=1}^{B} p_{\text{GT},i} \right) / B
}{
\sum_{i=1}^{B} p_i^2 + \sum_{i=1}^{B} p_{\text{GT},i}^2 - 2 \sum_{i=1}^{B} p_i p_{\text{GT},i} / B
}.
\end{equation}
Therefore, the loss $\mathcal{L}_{\text{total}}$ combining the strengths of both BCE and Kappa is now differentiable and hence implementable.

\subsection{Non-Quantum Implementation for IWD} \label{subsec: nonquantum}
%Number of parameters:
%QNN: 448
%2D CNN: 432
%depthwise convolution
To broaden accessibility for users of classical computers, we also provide the non-quantum counterpart of IWD-QUEEN, which is called IWD-Transformer, thereby
increasing the impact of this work.
IWD-Transformer maintains the preferable end-to-end architecture just as IWD-QUEEN, including the use of DAT for DDM encoding and an MLP for final feature classification.
However, the DAT output (i.e., $\bt_\text{DDM}$) does not rely on QUEEN-based QFRB for the subsequent feature extraction. 
Instead, $\bt_\text{DDM}$ is reshaped into a 3D tensor still with 16 channels, and then processed by an equally lightweight classical feature extraction block (CFEB), which consists of 2D convolution layers and GELU activation function.

Given the outstanding performance of IWD-QUEEN, we try to mimic its architectural design.
For example, we adopt depthwise convolution \cite{howard2017mobilenets}, which serves to emulate the non-shared-weight mechanism inherent in QFRB (i.e., quantum multihead mechanism), thereby preserving the conceptual design consistency between the two quantum and non-quantum architectures.
Note that depthwise convolution is a non-quantum scheme doing feature extraction independently among the 16 channels in CFEB.
For another example, just like the QFRB, which measures only two qubits per quantum circuit to produce a two-dimensional output, the CFEB outputs only a single scalar value for each channel.
Accordingly, the input dimension of the next MLP layer is simply adjusted to match the output dimension of CFEB.
Furthermore, CFEB is designed with a parameter size roughly comparable to that of QFRB, ensuring a comparable transfer from quantum design to its classical counterpart. 

Experimental results will demonstrate that IWD-Transformer still achieves outstanding detection performance, although IWD-QUEEN is even superior (cf. Section \ref{subsec: qq_analysis}). 
Under the design consistency established above, the experimental findings highlight the potential of quantum-inspired architecture in waterbody detection (so does its non-quantum implementation), underscoring the promising advantages of the proposed quantum and non-quantum techniques in related remote sensing applications.

\section{Experimental Results and Case Studies}\label{sec: experimental results}

This section comprehensively evaluates the two proposed algorithms, i.e., IWD-QUEEN and IWD-Transformer, to validate their effectiveness.
The experimental settings and the data descriptions are provided in Section \ref{subsec: experimental setting}.
Section \ref{subsec: qq_analysis} reports detailed comparisons with existing machine learning-based baselines, as well as a benchmark waterbody dataset.
An ablation study is conducted in Section \ref{subsec: ablation} to analyze the contribution of key components in IWD-QUEEN.
Through the evaluations in these two sections, we confirm that the proposed IWD-QUEEN demonstrates both robustness and accuracy in detecting inland waterbodies. 
Building on this foundation, Section \ref{subsec: case study} presents several case studies to further illustrate the practical applicability of IWD-QUEEN under real-world conditions, including the seasonal
variations, human activities, and the 2025 seismic event, etc.

\subsection{Experimental Settings} \label{subsec: experimental setting}

In this section, we will detail the ground truth (GT) and the CYGNSS DDM used for the experiment.
We adopt the MERIT Hydro global hydrography dataset \cite{MERIT_Hydro} as GT, which combines the MERIT DEM \cite{MERIT_DEM} with multiple waterbody datasets, including Global 1sec Water Body Map (G1WBM) \cite{G1WBM}, Global Surface Water (GSW) Occurrence \cite{GSW}, and OpenStreetMap \cite{OSM}.
MERIT Hydro is processed using the Google Earth Engine (GEE) platform\footnote{MERIT Hydro global hydrography dataset on GEE: \url{https://developers.google.com/earth-engine/datasets/catalog/MERIT_Hydro_v1_0_1}.}, and the ``viswth" band is used to represent river widths.
For visualization, river widths ranging from 1 m to 100 m are linearly mapped to grayscale, while those exceeding 100 m are assigned to 1. 
Subsequently, Otsu’s method \cite{Otsu} is applied to generate a binary water mask, owing to its robustness in thresholding and binarization.

For the DDMs used in the experiment, the Amazon River Basin in South America (62\textdegree W to 72\textdegree W, 2\textdegree S to 10\textdegree S) is selected as the region of interest (ROI). 
Training data is collected from January, April, July, and October of 2021, sampling one day per week.
This sampling strategy enables the network to capture seasonal variability across a nearly full-year time span while maintaining a reasonable training time.
Besides, two of the proposed methods are trained with a batch size of 100 for a total of 150 epochs, and we use the ADAM optimizer \cite{ADMMADAM} with a learning rate set to 0.001, and the loss function follows the formulation introduced in Section \ref{subsec: loss}.
The testing data are constructed using DDMs from the entire year of 2023.
Compared to the seasonal sampled training data, the full-year testing data enables evaluation of the model’s generalization at different times of the year. 

To ensure DDM quality and prevent the model from learning inaccurate representations, a series of filtering steps is applied to both training and testing data.
First, DDMs exhibiting known signal anomalies caused by hardware malfunctions or environmental interference are excluded based on quality flags provided by CYGNSS. 
In addition, DDMs collected under suboptimal configurations are removed, such as incidence angles exceeding 65$^\circ$ or receiver antenna gains below 0 dB \cite{HydroQUEEN}, as these conditions are known to distort the observed scattering signatures. 
To further ensure reliability, DDMs with signal-to-noise ratios (SNRs) below 2 dB are discarded \cite{rajabi2020evaluation}, to retain those DDMs with sufficiently clear surface scattering patterns. 
The SNR is computed as follows:
\begin{equation} 
\text{SNR}= {10} \log\left(\frac{X_{\text{max}}}{N_{\text{avg}}}\right), 
\end{equation} 
where $X_{\text{max}}$ denotes the maximum value in the DDM $\bX$, and $N_{\text{avg}}$ represents the average per-bin noise count of $\bX$ \cite{CYGNSS_Handbook}. 
All the aforementioned filtering attributes, as well as the DDM data, are directly available from the CYGNSS Level 1 science data record dataset and can be downloaded via the Earthdata Search platform\footnote{CYGNSS Level 1 science data record dataset on Earthdata Search: \url{https://search.earthdata.nasa.gov/search}.}.
All experiments are conducted on an equipment with an Intel i9-13900 CPU and an NVIDIA GeForce RTX 4090 GPU.
All methods are implemented on PyTorch 2.5.1, PennyLane 0.37.0 \cite{Pennylane}, and MATLAB R2024a.

\begin{figure*}
    \centering
    \includegraphics[width=1\linewidth]{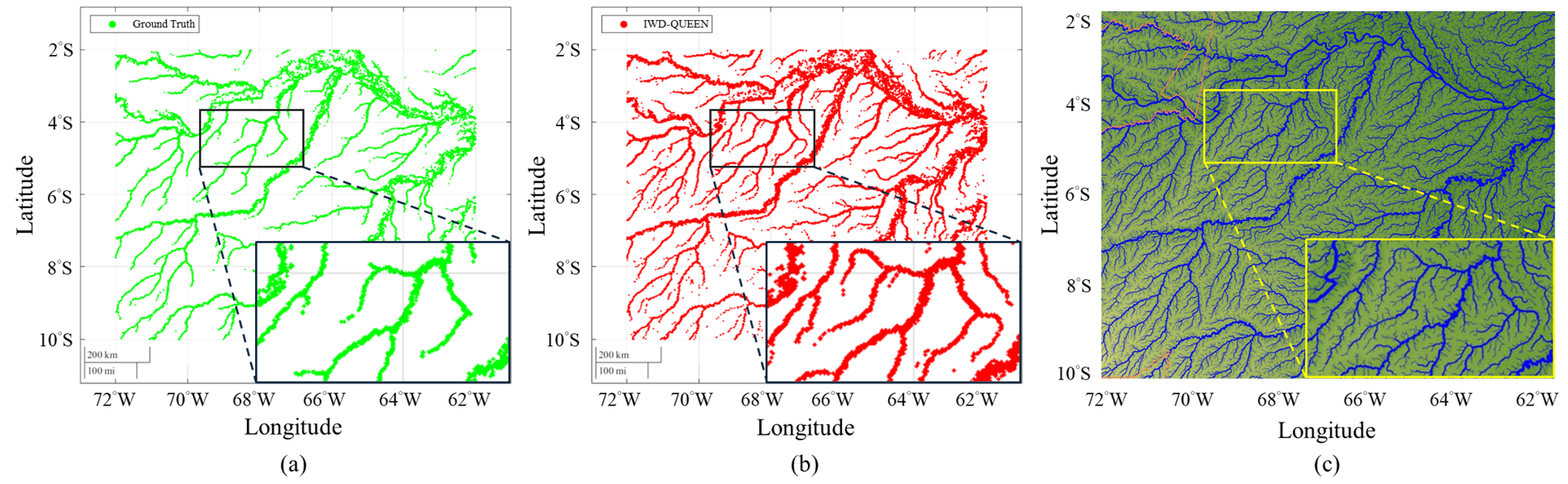}
    \caption{Visual comparison of detection results in Amazon River Basin. 
    (a) Ground truth (GT). 
    (b) Detection results by IWD-QUEEN. 
    (c) The image depicted using NASA’s Shuttle Radar Topography Mission (SRTM) and the river networks are shown in blue.
    The zoomed-in regions highlight that IWD-QUEEN can detect small tributaries and headwaters compared to GT, demonstrating its capability in identifying inland waterbodies.}
    \label{fig: finer_than_GT}
\end{figure*}

\begin{figure*}
    \centering
    \includegraphics[width=1\linewidth]{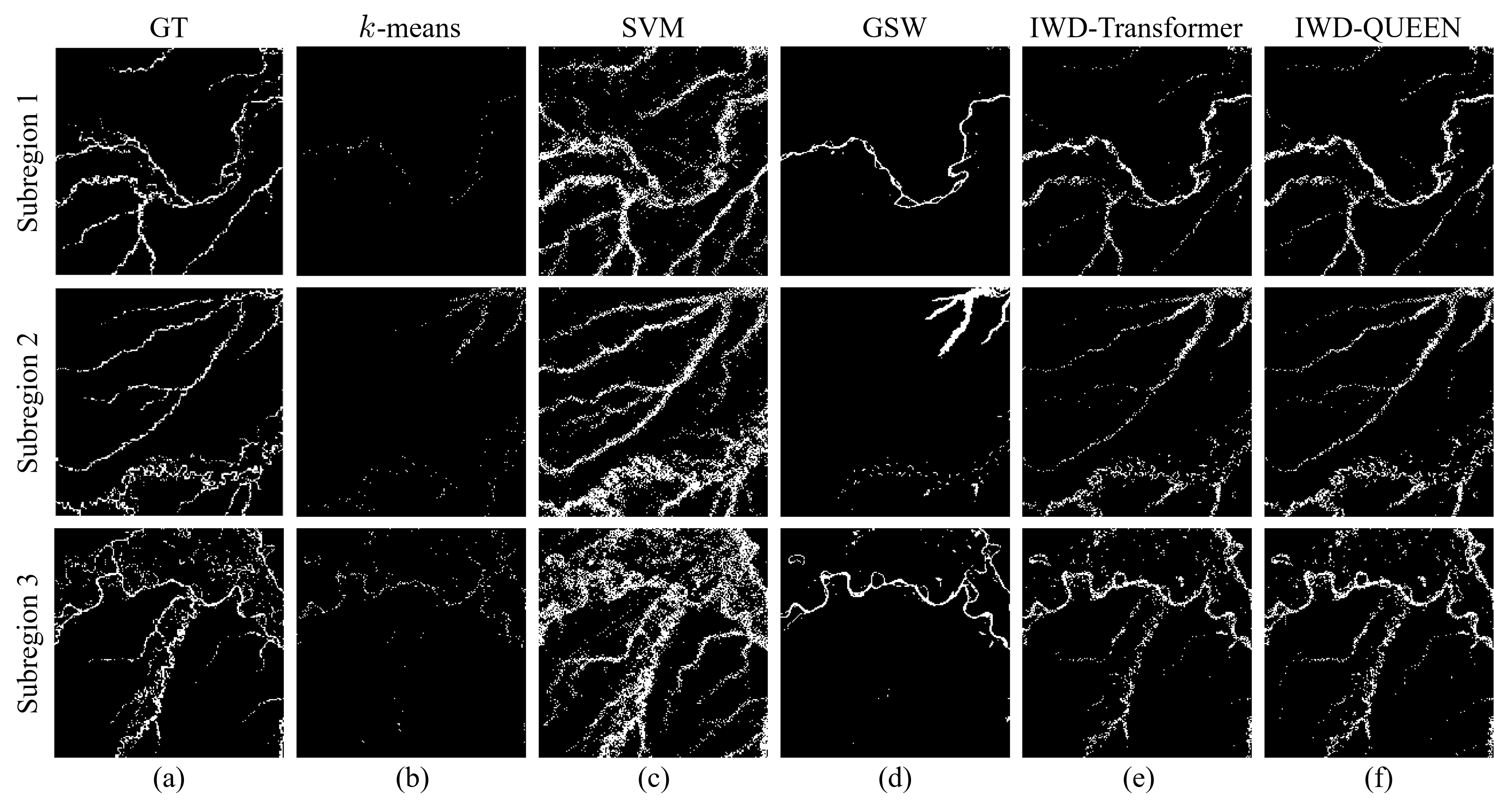}
    \caption{Qualitative study on each subregion. 
    (a) Ground Truth. 
    The detection results obtained by (b) $k$-means, (c) SVM, (d) GSW, (e) IWD-Transformer, and (f) IWD-QUEEN.
    }
    \label{fig: Qualitative comparison}
\end{figure*}

\subsection{Qualitative and Quantitative Analysis} \label{subsec: qq_analysis}

To assess the performance of the proposed IWD-QUEEN and IWD-Transformer, we conduct a comparative analysis against $k$-means clustering \cite{kmeans}, support vector machine (SVM) \cite{SVM}, and the GSW product \cite{GSW} as benchmarks. 
$k$-means clustering \cite{kmeans} is an unsupervised machine learning algorithm that partitions samples into $k$ clusters based on feature similarity (commonly using the square of the Euclidean distance).
Since the IWD problem is inherently a binary classification task, $k$ is intuitively set to $2$.
SVM is a supervised learning algorithm that constructs an optimal decision boundary, known as a hyperplane, to separate samples from different classes with maximum margin.
Besides, SVM leverages a kernel function (radial basis function (RBF) is used in this article) to project the samples into a higher-dimensional feature space, making separation feasible.
The Global Surface Water (GSW) dataset is a satellite-derived product generated by analyzing over three million Landsat satellite images \cite{Landsat} collected from 1984 to 2015. 
Due to its long-term coverage and public availability, GSW has become a widely used reference for evaluating climate impacts, wetland state changes, and water resource management. 

For the qualitative analysis, we first observe that IWD-QUEEN is capable of capturing finer-scale inland waterbodies (cf. Figure \ref{fig: finer_than_GT}(b)), especially narrow tributaries and headwaters, which are often missing in GT (cf. Figure \ref{fig: finer_than_GT}(a)). 
Moreover, the detection result of IWD-QUEEN show a closer resemblance to the image depicted using NASA’s Shuttle Radar Topography Mission (SRTM)\footnote{For more information, see: \url{https://earthobservatory.nasa.gov/images/7823/source-of-the-amazon-river}.} (cf. Figure \ref{fig: finer_than_GT}(c)), and the river networks are shown in blue.
To ensure a meaningful comparison, we avoid the above situation by selecting three representative subregions from ROI, each exhibiting different river network densities: sparse (69\textdegree W to 71\textdegree W, 3\textdegree S to 5\textdegree S), moderate (63\textdegree W to 65\textdegree W, 4\textdegree S to 6\textdegree S), and dense (65\textdegree W to 67\textdegree W, 2\textdegree S to 4\textdegree S), as shown in Figure \ref{fig: Qualitative comparison}(a).
However, due to the different spatial resolutions and temporal coverage of the CYGNSS DDMs, GSW, and GT, the ROI is uniformly partitioned into grids of size 0.01$^\circ \times$ 0.01$^\circ$. 
This grid-based representation allows for fair comparison and enables pixel-wise evaluation of IWD performance. 
In particular, within each grid, sample values are aggregated via averaging, and a threshold of $0.5$ is subsequently applied to binarize the results for classification. 

We further conduct qualitative analysis based on the detection results over three representative subregions, as illustrated in Figure \ref{fig: Qualitative comparison}.
Figure \ref{fig: Qualitative comparison}(b) shows that the detection result of the $k$-means clustering fails to effectively identify inland waterbodies across all subregions.
The results reveal that almost no meaningful river structure is preserved.
This can be attributed to the inherent limitation of $k$-means clustering, which relies on Euclidean distance in a linear feature space and therefore struggles to capture the complex patterns present in CYGNSS DDMs. 
Compared to $k$-means clustering, SVM demonstrates improved coverage of river networks, as can be seen from Figure \ref{fig: Qualitative comparison}(c).
However, it also introduces several notable issues. 
First, the results suffer from overdetection, with the detected waterbodies extending beyond the GT.
Second, we observe the presence of considerable noise patterns, particularly in subregions 1 and 3.
Although SVM incorporates a kernel method to project data into a higher-dimensional feature space, one fundamental limitation of SVM lies in its inability to capture the contextual relationships within the DDMs.
The GSW product is displayed in Figure \ref{fig: Qualitative comparison}(d).
While it successfully delineates major rivers with clear boundaries, a considerable portion of upstream channels and tributaries is omitted.
This omission is primarily attributed to the inherent limitations of optical remote sensing \cite{HydroQUEEN}, which is sensitive to cloud coverage and dense vegetation, leading to underdetection in forested or persistently shaded regions.
In contrast, the performance of IWD-Transformer and IWD-QUEEN is significantly improved, as shown in Figure \ref{fig: Qualitative comparison}(e) and \ref{fig: Qualitative comparison}(f).
The detected waterbodies exhibit better spatial continuity and alignment with river patterns in GT.
Moreover, IWD-QUEEN achieves even better recovery of the entire river network and successfully captures many finer-scale river features that appear less prominent in IWD-Transformer (cf. lower-left part in subregion 1 and middle-left part in subregion 3 in Figure \ref{fig: Qualitative comparison}(f)). 
This reflects the enhanced expressivity and robustness of IWD-QUEEN, achieved through quantum feature refinement. 
 
\begin{table}[t]
    \caption{Quantitative performance of the proposed methods and baselines over three subregions of the Amazon River Basin.
    Higher values across all the metrics indicate better results.}
    \begin{center}
        \scalebox{0.85}{
            \begin{tabular}{c|c|c|c|c|c|c}
                \hline\hline
                \multirow{2}{*}{} & \multirow{2}{*}{Methods} & \multicolumn{5}{c}{\rule{0pt}{2ex}Metrics \rule{0pt}{2ex}} \\
                \cline{3-7}
                & & Recall  &Precision &F1 &OA &$\kappa$\\
                \hline
                \multirow{5}{*}{\shortstack{Subregion 1\\(sparse)}} 
                &$k$-means \cite{kmeans} &0.02 &\textbf{0.57} &0.03 &\textbf{0.95} &0.03 \\
                &SVM \cite{SVM} &\underline{\textbf{0.72}} &0.27 &0.39 &0.90 &0.34 \\
                &GSW \cite{GSW} &0.26 &\underline{\textbf{0.66}} &0.38 &\underline{\textbf{0.96}}  &0.36 \\
                &IWD-Transformer &0.42 &0.52 &\textbf{0.46} &\underline{\textbf{0.96}} &\textbf{0.44} \\
                &IWD-QUEEN &\textbf{0.50} &0.49 &\underline{\textbf{0.49}} &\textbf{0.95} &\underline{\textbf{0.47}} \\
                \hline
                
                \multirow{5}{*}{\shortstack{Subregion 2\\(moderate)}} 
                &$k$-means \cite{kmeans} &0.06 &\textbf{0.43} &0.10 &\underline{\textbf{0.95}} &0.09 \\
                &SVM \cite{SVM} &\underline{\textbf{0.72}} &0.26 &0.39 &0.88 &0.33 \\
                &GSW \cite{GSW} &0.21 &0.42 &0.28 &\textbf{0.94} &0.25 \\
                &IWD-Transformer &0.36 &\underline{\textbf{0.44}} &\textbf{0.40} &\textbf{0.94} &\textbf{0.37} \\
                &IWD-QUEEN &\textbf{0.45} &\underline{\textbf{0.44}} &\underline{\textbf{0.44}} &\textbf{0.94} &\underline{\textbf{0.41}}  \\
                \hline

                \multirow{5}{*}{\shortstack{Subregion 3\\(dense)}} 
                &$k$-means \cite{kmeans} &0.07 &\textbf{0.47} &0.12 &\underline{\textbf{0.94}} &0.10 \\
                &SVM \cite{SVM} &\underline{\textbf{0.68}} &0.24 &0.36 &0.85 &0.29 \\
                &GSW \cite{GSW} &0.28 &\underline{\textbf{0.58}} &0.38 &\underline{\textbf{0.94}} &0.35 \\
                &IWD-Transformer &0.38 &0.45 &\textbf{0.42} &\textbf{0.93} &\textbf{0.38} \\
                &IWD-QUEEN &\textbf{0.45} &0.43 &\underline{\textbf{0.44}} &\textbf{0.93} &\underline{\textbf{0.40}} \\
                \hline\hline
            \end{tabular}
        }
    \end{center}
    \label{tab: Quantitative}
\end{table}

To analyze the performance of our method further, we need quantitative metrics.
Specifically, we adopt Recall, Precision, F1-score (F1), overall accuracy (OA), and $\kappa$ \cite{QUEEN-G} for evaluation.
Higher values across all the metrics indicate better performance of the method.
To facilitate the interpretation of the evaluation metrics, we first introduce the standard classification terms: TP, FP, TN, and FN, referring to true positives, false positives, true negatives, and false negatives, respectively.
In this article, TP refers to the samples that are inland waterbodies and are correctly identified; FP denotes the samples that are land surfaces but are incorrectly predicted as inland waterbodies; TN indicates the correctly predicted land surface samples; FN refers to inland waterbodies that are mistakenly classified as land surfaces.
Based on the above illustration, Recall and Precision can be defined as 
\begin{align} 
\text{Recall} &= \frac{\text{TP}}{\text{TP} + \text{FN}}
\\
\quad \text{Precision} &= \frac{\text{TP}}{\text{TP} + \text{FP}},
\end{align}
where Recall evaluates how well the algorithm recovers the actual inland waterbodies from GT, and Precision assesses how many predicted waterbodies are truly correct. 
We remark that Recall and Precision reflect complementary aspects of classification performance and should be considered jointly, as focusing on only one may lead to misleading conclusions. 
Therefore, F1 is adopted as it calculates the harmonic mean of Recall and Precision to achieve a comprehensive evaluation, which is defined as
\begin{equation} 
\text{F1} = 2 \times \frac{\text{Precision} \times \text{Recall}}{\text{Precision} + \text{Recall}}.
\end{equation}
OA calculates the proportion of correctly classified samples over the detection results, and is expressed as
\begin{equation} 
\text{OA} = \frac{\text{TP} + \text{TN}}{\text{TP} + \text{TN} + \text{FP} + \text{FN}}.
\end{equation}
However, OA may be insufficient when class distributions are imbalanced.
To address this limitation, $\kappa$ is also adopted, as it accounts for the interrater reliability, thereby offering a more reliable evaluation than OA.
To derive $\kappa$, we first define the hypothetical probability of chance agreement, denoted as $\text{P}_e$, as follows
\begin{equation}
\text{P}_e = \frac{(\text{TP} + \text{FP})(\text{TP} + \text{FN}) + (\text{FN} + \text{TN})(\text{FP} + \text{TN})}{(\text{TP} + \text{TN} + \text{FP} + \text{FN})^2}.
\end{equation}
Then, the $\kappa$ can be derived as 
\begin{equation}
\kappa = \frac{\text{OA} - \text{P}_e}{1 - \text{P}_e}.
\end{equation}

We analyze the quantitative results in Table \ref{tab: Quantitative} with the visual comparisons shown in Figure \ref{fig: Qualitative comparison}, focusing on the trade-offs between different metrics and their relationship with the detection performance.
For each metric, the best performance is indicated by boldface with underlining, while the second-best is shown in boldface only.
First, although SVM exhibits the highest Recall across all subregions, its Precision is significantly lower than that of the other methods.
This imbalance is visually supported by Figure \ref{fig: Qualitative comparison}(c), where SVM classifies most samples as waterbodies.
As a result, many FP are counted, leading to overdetection and reducing the detection reliability.
In contrast, $k$-means clustering and GSW show relatively high Precision in subregions 1 and 3, indicating a careful prediction behavior.
This conservative nature can also be seen from Figures \ref{fig: Qualitative comparison}(b) and \ref{fig: Qualitative comparison}(d), where both methods detect only main river structures while omitting finer tributaries and headwaters.
However, this cautious strategy leads to substantial FN, resulting in low Recall.
Given the complementary roles of Recall and Precision, the F1-score serves as a more balanced evaluation metric.
Notably, IWD-Transformer and IWD-QUEEN rank among the top two across all subregions regarding F1, indicating that our proposed methods achieve a better balance between detecting true waterbodies and avoiding false alarms. 

Although OA remains high for all methods (usually above 0.90), its discrimination is limited because the number of land surface samples vastly exceeds that of waterbody samples.
This makes OA susceptible to the dominant class (i.e., land surface).
For instance, even $k$-means, which fails to detect most waterbodies, still achieves an OA of 0.95 in subregion 1.
Unlike OA, $\kappa$ accounts for the possibility of agreement occurring by chance, offering a more balanced view of classification performance under imbalanced conditions. 
The proposed IWD-QUEEN and IWD-Transformer consistently rank among the top two in terms of $\kappa$ across all subregions, clearly demonstrating the robustness and reliability of our methods in accurately detecting inland waterbodies.
Remarkably, IWD-QUEEN, together with its parallel quantum multihead scheme, works in a near-real-time manner (i.e., 6E-3 seconds per DDM).
These results highlight the practicality of our framework, paving the way for timely IWD.

\subsection{Ablation Study} \label{subsec: ablation}
\begin{table}[t]
    \caption{Ablation study of three subregions, along with their average for the proposed IWD-QUEEN, where the marker ``\CheckmarkBold" means the corresponding module is used.
    Higher values across all the metrics indicate better results.}
    \begin{center}
        \scalebox{0.85}{
            \begin{tabular}{c|c c|c|c|c|c|c}
                \hline\hline
                \multirow{3}{*}{} & \multicolumn{2}{c|}{Module} & \multicolumn{5}{c}{Metrics} \\
                \cline{2-3} \cline{4-8}
                & SE & QC & Recall & Precision & F1 & OA & $\kappa$ \\
                
                \hline
                \multirow{3}{*}{\shortstack{Subregion 1\\(sparse)}} 
                &               &                & 0.42 & \underline{\textbf{0.52}} & 0.46 & \underline{\textbf{0.96}} & 0.44 \\
                &               &\CheckmarkBold  & 0.44 & 0.50 & 0.47 & 0.95 & 0.44 \\
                &\CheckmarkBold &\CheckmarkBold  & \underline{\textbf{0.50}} & 0.49 & \underline{\textbf{0.49}} & 0.95 & \underline{\textbf{0.47}} \\

                \hline
                \multirow{3}{*}{\shortstack{Subregion 2\\(moderate)}} 
                &               &                & 0.36 & \underline{\textbf{0.44}} & 0.40 & \underline{\textbf{0.94}} & 0.37 \\
                &               &\CheckmarkBold  & 0.38 & \underline{\textbf{0.44}} & 0.41 & \underline{\textbf{0.94}} & 0.38 \\
                &\CheckmarkBold &\CheckmarkBold  & \underline{\textbf{0.45}} & \underline{\textbf{0.44}} & \underline{\textbf{0.44}} & \underline{\textbf{0.94}} & \underline{\textbf{0.41}} \\

                \hline
                \multirow{3}{*}{\shortstack{Subregion 3\\(dense)}} 
                &               &                & 0.38 & \underline{\textbf{0.45}} & 0.42 & \underline{\textbf{0.93}} & 0.38 \\
                &               &\CheckmarkBold  & 0.39 & 0.44 & 0.42 & \underline{\textbf{0.93}} & 0.38 \\
                &\CheckmarkBold &\CheckmarkBold  & \underline{\textbf{0.45}} & 0.43 & \underline{\textbf{0.44}} & \underline{\textbf{0.93}} & \underline{\textbf{0.40}} \\

                \hline\hline
                \multirow{3}{*}{Average} 
                &               &                & 0.39 & \underline{\textbf{0.47}} & 0.43 & \underline{\textbf{0.94}} & 0.40 \\
                &               &\CheckmarkBold  & 0.40 & 0.46 & 0.43 & \underline{\textbf{0.94}} & 0.40 \\
                &\CheckmarkBold &\CheckmarkBold  & \underline{\textbf{0.47}} & 0.45 & \underline{\textbf{0.46}} & \underline{\textbf{0.94}} & \underline{\textbf{0.43}} \\

                \hline\hline
            \end{tabular}
        }
    \label{tab: ablation study}
    \end{center}
\end{table}

In this experiment, we conduct an ablation study to investigate the influence of two essential quantum components of IWD-QUEEN.
The first component is the SE module described in \eqref{QFRB}.
We will demonstrate that the SE module is better than a shallow entanglement strategy (refer to \cite{HyperQUEEN}) for capturing quantum feature correlations.
The second component is the quantum computing (QC) module, which is evaluated using the QFRB in Section \ref{subsec: QFRB}.
Alternatively, we replace the QC module with a classical counterpart (cf. Section \ref{subsec: nonquantum}) that structurally emulates the quantum multihead scheme.
In the absence of the QC module, the input dimension of the subsequent MLP layer is accordingly adjusted to 16.
Furthermore, the parameter size of the classical module is designed to be roughly comparable to the QC module to ensure a fair comparison.

We test the proposed IWD-QUEEN by evaluating three configurations across all three subregions introduced in Section \ref{subsec: qq_analysis}.
We remark that the SE module is inherently tied to the QUEEN and thus depends on the presence of the QC module.
Consequently, when the QC module is removed, the SE module is also naturally excluded, resulting in only three valid configurations for comparison.
The ablation study results for the three subregions and their average are quantitatively displayed in Table \ref{tab: ablation study}.
We observe that the performance of IWD-QUEEN consistently improves as each quantum component is added.
In particular, enabling the QC module alone yields slight gains in F1 and $\kappa$ across all subregions, while incorporating both SE and QC modules leads to the best overall detection performance.
This highlights the critical role of quantum refinement and strong entanglement in enhancing feature representation and improving classification performance for IWD.

\begin{figure} [t]
    \centering
    \includegraphics[width=0.9\linewidth]{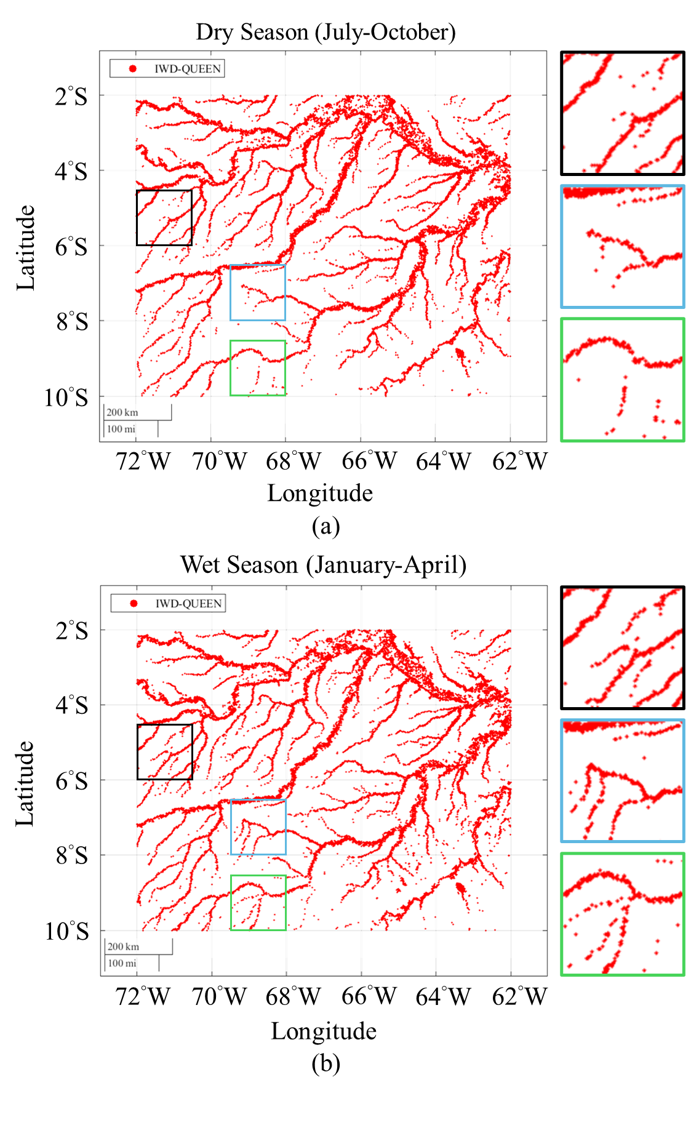}
    \caption{Detection results obtained by IWD-QUEEN over Amazon River Basin during (a) dry season and (b) wet season.}
    \label{fig: amazon season}
\end{figure}

\subsection{Case Studies Under Real-World Conditions} \label{subsec: case study}

As demonstrated in Section \ref{subsec: qq_analysis}, IWD-QUEEN exhibits robust detection capabilities, supported by multiple evaluation metrics.
In this section, we present the surface water detection results using IWD-QUEEN, and further assess its performance across selected regions under varying seasonal conditions.
In addition to the Amazon River Basin in South America, which is used as the ROI in Section \ref{subsec: experimental setting}, we also incorporate the Mekong River Basin in Southeast Asia (101\textdegree E to 105\textdegree E, 16\textdegree N to 20\textdegree N) for model training and testing. 
Besides, the training and testing data for the Mekong River Basin follow the same temporal setting described in Section \ref{subsec: experimental setting}.
This section also investigates the model's applicability and stability under diverse conditions, including seasonal variations in precipitation (e.g., dry season and wet season) and agricultural activities (e.g., planting season and harvesting season). 
Finally, we examine the response of IWD-QUEEN to a recent major seismic event in 2025, evaluating its ability to detect earthquake-induced geomorphological changes, such as landslides, in remote mountainous areas where conventional disaster information is often difficult to acquire.

\subsubsection{Amazon River Basin Analysis} \label{subsubsec: case study Amazon}

During the training phase using DDMs from the Amazon River Basin, we intentionally selected inland areas in order to enhance the contrast between the DDM signatures of dense jungle canopy and river water surfaces. 
This approach also helps to minimize detection errors that may arise due to the shallow waters commonly found near coastal regions.
Precipitation variability in this region is subdivided into two distinct seasons based on the NASA MERRA-2 reanalysis dataset\footnote{The NASA MERRA-2 reanalysis dataset: \url{https://gmao.gsfc.nasa.gov/reanalysis/MERRA-2/}.}: the dry season (July to October) and the wet season (January to April). 
To more clearly highlight the seasonal differences, data from the transitional months (e.g., May to June) are excluded.

As illustrated in Figure \ref{fig: amazon season}, the number of observable tributaries in the Amazon region increases markedly during the wet season. 
To facilitate a more detailed analysis, three specific tributary regions are selected for further discussion.
\begin{itemize}
    \item Black Box Region (70.5\textdegree W to 72\textdegree W, 4.5\textdegree S to 6\textdegree S):\\
    This area includes both the middle and upper reaches of one of the tributaries of the Amazon River.
    During the dry season, the river flow lines are scattered and discontinuous; however, during the wet season, the tributary system shown here forming a continuous distribution of waterbody detections and presenting a complete waterbody structure.
    This indicates that, in Amazon River Basin, the wet season is the critical period for observing the full extent of the tributary.

    \item Blue Box Region (68\textdegree W to 69.5\textdegree W, 6.5\textdegree S to 8\textdegree S):\\
    The northern part of this region is the main river channel, with minimal seasonal change between dry and wet seasons. 
    However, the southern part is the upstream tributary area, where three northward flowing tributaries appear in the wet season, covering the lower half of the region, showing significant waterbody expansion.

    \item Green Box Region (68\textdegree W to 69.5\textdegree W, 8.5\textdegree S to 10\textdegree S):\\
    This region is located south of the main river channel, and the wet season reveals a watershed formed by three tributaries from the south. 
    This demonstrates that the IWD-QUEEN is highly sensitive to weak river signals, effectively revealing the expansion of river networks in deep forest areas.
\end{itemize}

\begin{table}[t]
\caption{Detection rates of the selected two seasons in the defined box areas.}
\centering
\begin{tabular}{l|c|c}
\hline\hline
& Dry Season & Wet Season \\
\hline
Black Box & 715 / 3483 = 20.5\% & 895 / 4355 = 20.6\% \\
\hline
Blue Box  & 756 / 3118 = 23.6\% & 930 / 4625 = 20.1\% \\
\hline
Green Box & 313 / 5449 = 5.7\%  & 462 / 7607 = 6\%    \\
\hline
\hline
\end{tabular} \label{tab: amazon season}
\end{table}

To quantify the differences between the results in dry and wet seasons, we list the detection rates (the number of detected points divided by the total observations) of the three box regions for reference.
In the black box, the detection rates during the dry/wet seasons (cf. Table \ref{tab: amazon season}) are 20.5\%/20.6\%.
In the blue box, they are 23.6\%/20.1\%.
In the green box, they are 5.7\%/6\%.
These results show that the detection rates during the dry and wet seasons are similar, indicating that the river width of the Amazon region does not show significant changes with precipitation.
However, more tributaries are detected during the wet season.

\begin{figure} [t]
    \centering
    \includegraphics[width=0.9\linewidth]{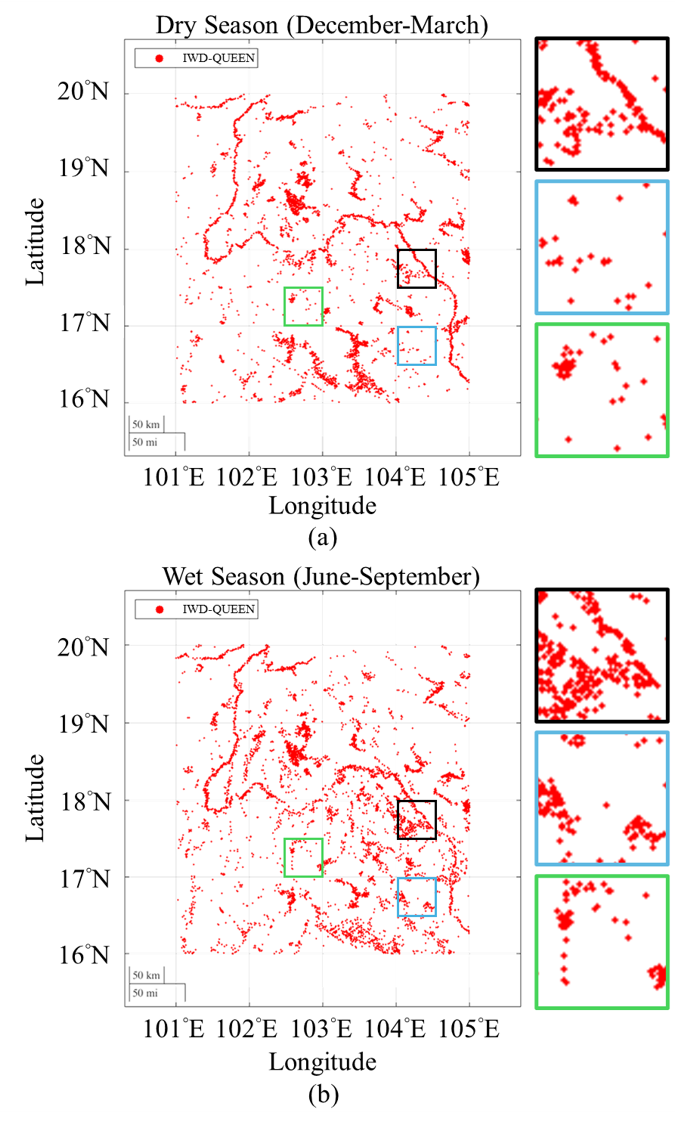}
    \caption{Detection results obtained by IWD-QUEEN over Mekong River Basin during (a) dry season and (b) wet season.}
    \label{fig: mekong season}
\end{figure}

\subsubsection{Mekong River Basin Analysis} \label{subsubsec: case study mekong}

On the other hand, for the Mekong River Basin, the decision to select inland areas for training is made to avoid the extensive rice paddies near the Mekong Delta.
According to the precipitation record in NASA MERRA-2, the dry season in Thailand is defined from December to March, while the wet season spans from June to September.
The detection results presented in Figure \ref{fig: mekong season} show that the Mekong River Basin is more complex than the Amazon River Basin.
In addition to the main channel of the Mekong River, which flows south from  101.7\textdegree E, 19.5\textdegree N and turns southeast at 104\textdegree E, several scattered points and larger detection zones are observed in the surrounding areas.

To analyze the seasonal differences, we selected three specific regions for comparative analysis between the two seasons.
\begin{itemize}
    \item Black Box Region (104\textdegree E to 104.5\textdegree E, 17.5\textdegree N to 18\textdegree N):\\
    In this region, the Mekong River flows from the north-eastward, and the lower half consists of the Songkhram River Basin (104.2\textdegree E, 17.7\textdegree N) and surrounding wetlands. 
    During the wet season, the wetland area expands significantly, better describing its extent.
    The Mekong River channel in this region also becomes noticeably wider.

    \item Blue Box Region (104\textdegree E to 104.5\textdegree E, 16.5\textdegree N to 17\textdegree N):\\
    During the dry season, only scattered detection points are observed, with no significant waterbody signals. 
    However, during the wet season, small lake systems such as Lake Nam Phung (104.08\textdegree E, 16.5\textdegree N) become more pronounced, showing that small waterbodies are more detectable in the wet season.

    \item Green Box Region (102.5\textdegree E to 103\textdegree E, 17\textdegree N to 17.5\textdegree N):\\
    This area includes Lakes Huai Luang (102.6\textdegree E, 17.3\textdegree N) and Han Kumphawapi (103\textdegree E, 17.2\textdegree N).
    During the dry season, only Huai Luang is detected, but during the wet season, both lakes show clear area-like wetland/lake signals.
\end{itemize}

\begin{table}[t]
\caption{Detection rates of the selected two seasons in the defined box areas.}
\centering
\begin{tabular}{l|c|c}
\hline\hline
& Dry Season & Wet Season \\
\hline
Black Box & 172 / 3102 = 5.5\% & 313 / 3181 = 9.8\% \\
\hline
Blue Box  & 28 / 3292 = 0.9\% & 99 / 2513 = 3.9\% \\
\hline
Green Box & 53 / 3321 = 1.6\%  & 77 / 3543 = 2.2\%    \\
\hline
\hline
\end{tabular} \label{tab: mekong season}
\end{table}

The quantification of detection rates (cf. Table \ref{tab: mekong season}) shows: the detection rates for the black box during the dry/wet seasons are 5.5\%/9.8\%, for the blue box they are 0.9\%/3.9\%, and for the green box, they are 1.6\%/2.2\%.
These results show that the detection rates in the wet season are significantly higher than in the dry season, indicating that the river width and surface water in the Thailand region are more strongly influenced by seasonal rainfall, with the environmental resilience of water resources being notably weaker than in the Amazon River Basin.

\subsubsection{Surface Water Detection and Human Activity in the Mekong River Basin} \label{subsubsec: case study human}

\begin{figure} [ht]
    \centering
    \includegraphics[width=0.8\linewidth]{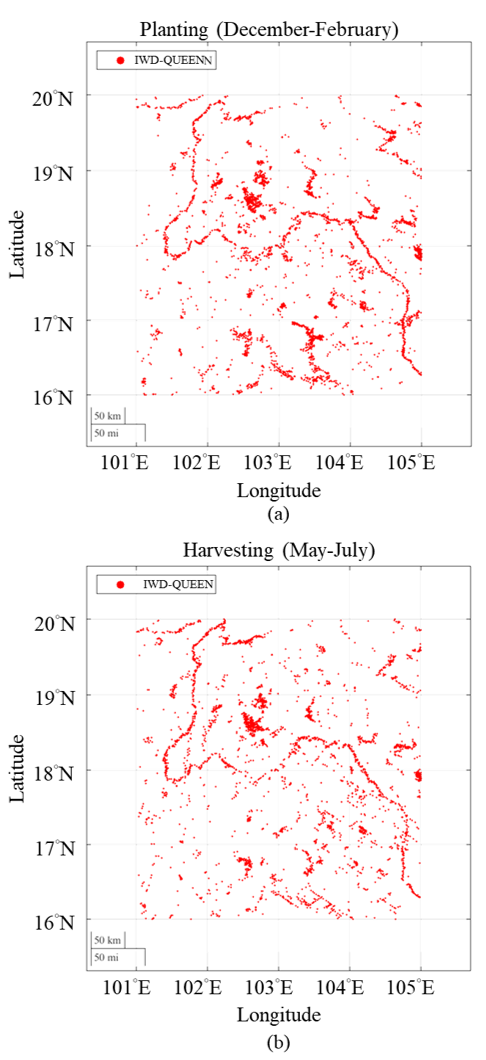}
    \caption{Detection results obtained by IWD-QUEEN over Mekong River Basin during (a) planting season and (b) harvesting season.}
    \label{fig: mekong agri}
\end{figure}

In Southeast Asia, the main agricultural sector requiring large amounts of water is irrigation, with rice farming being predominant.
Based on information from the Food and Agriculture Organization (FAO)\footnote{Food and Agriculture Organization (FAO): \url{https://www.fao.org/}.} of the United Nations, we define December to February as the planting season for rice in Thailand, and May to July as the harvesting season.
Rice farming requires significant water for paddy fields in the planting stages, while the demand for irrigation decreases later, and dense plant growth makes it difficult to detect water surfaces, particularly during the harvesting season.

Using Global Agricultural Monitoring System (GADAS)\footnote{Global Agricultural Monitoring System (GADAS): \url{https://geo.fas.usda.gov/GADAS/index.html}.} and ESA WorldCover Viewer\footnote{ESA WorldCover Viewer: \url{https://esa-worldcover.org/en}.}, we find that the southern part of the Mekong River (within Thailand territory) is primarily composed of rice paddies, while the northern part (in Laos territory) is more mountainous-like terrain.
As shown in Figure \ref{fig: mekong agri}, the waterbody changes on the northern bank of the Mekong River during the planting and harvesting seasons are less pronounced.
For example, Lake Nam Ngum at 102.5\textdegree E, 18.5\textdegree N shows no significant changes during these two seasons.
In contrast, on the southern bank of the Mekong River, waterbodies often show fragmented outlines, such as Lake Lam Pao at 103.5\textdegree E, 16.7\textdegree N, which is nearly invisible during the harvesting season.

\begin{figure*} [t]
    \centering
    \includegraphics[width=1\linewidth]{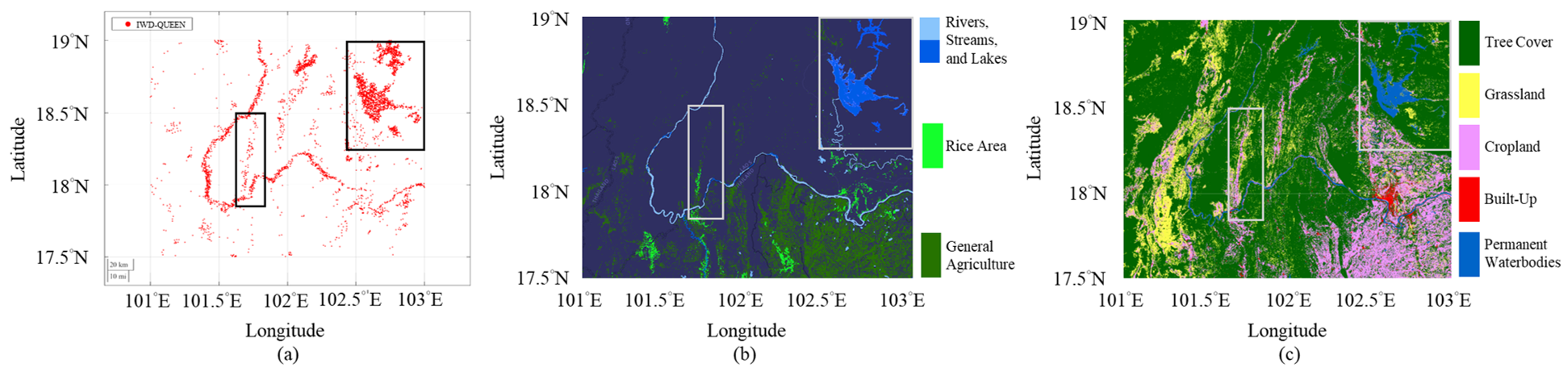}
    \caption{(a) Detection results obtained by IWD-QUEEN over Mekong River Basin.
    Maps of (b) GADAS and (c) WorldCover Viewer.}
    \label{fig: mekong human}
\end{figure*}

Notably, although the planting season corresponds to the dry season and the harvesting season corresponds to the wet season, waterbody detection is higher during the planting season, indicating that human agricultural activities have a more significant impact on surface water than precipitation.
Additionally, IWD-QUEEN not only delineates the detailed shapes of surface waterbodies, such as the Nam Ngum Reservoir located at 102.7\textdegree E, 18.5\textdegree N (cf. the upper-right box in Figure \ref{fig: mekong human}(a)) but also detects a north-south water extension at 101.7\textdegree E, 18.5\textdegree N (cf. the middle box in Figure \ref{fig: mekong human}(a)).
Upon cross-referencing with the GADAS and ESA WorldCover Viewer, we identify this area as a general agricultural zone, including rice cultivation (cf. Figures \ref{fig: mekong human}(b) and \ref{fig: mekong human}(c)).
This finding suggests that the developed IWD-QUEEN exhibits robust and sensitive detection capabilities for surface waterbodies.
This motivates us that in the following research, we will aim to further classify these waterbodies into natural systems and artificial ones, such as rice paddies and irrigation canals.
Such advancements could prove valuable for applications in civilizational surveys and agricultural economic strategies.

\begin{figure} [t]
    \centering
    \includegraphics[width=0.9\linewidth]{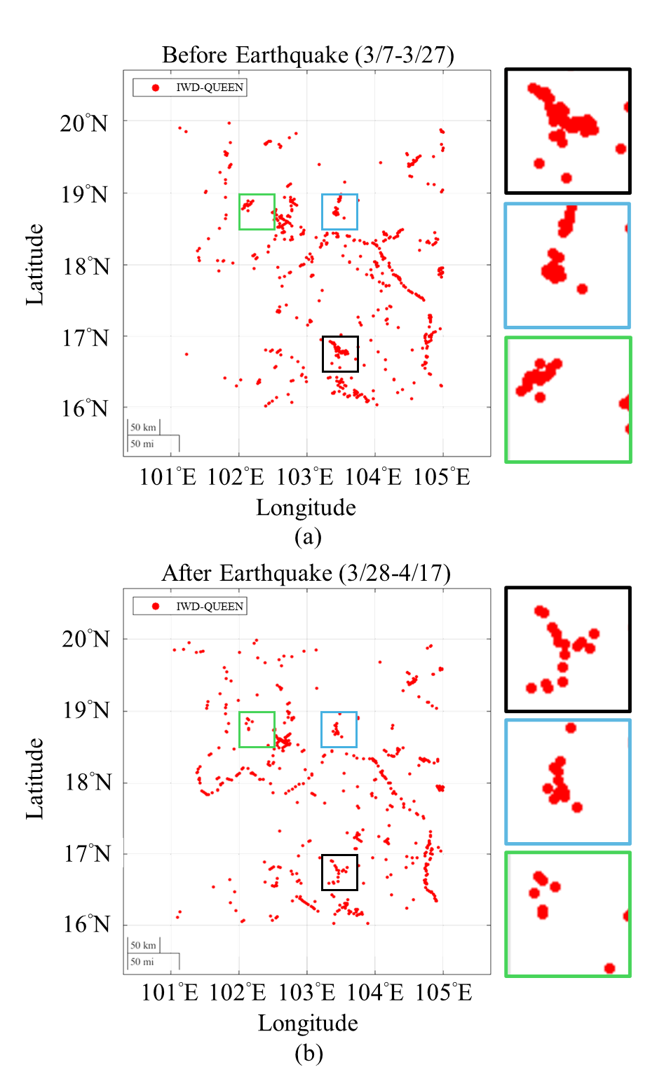}
    \caption{Detection results obtained by IWD-QUEEN over Mekong River Basin (a) before and (b) after the 2025 Myanmar earthquake.}
    \label{fig: mekong earthquake}
\end{figure}

\subsubsection{Impact of the 2025 Southeast Asia Mega-Earthquake Event on Surface Water Detection} \label{subsubsec: case study earthquake}

The 2025 Myanmar earthquake, with a magnitude of 7.7, caused significant geological changes, particularly in the region’s landscape. 
The seismic event, centered near Mandalay, triggered ground deformation and affected waterbodies by altering river courses and impacting water quality. 
The earthquake’s aftershocks further disrupted the landscape, resulting in landslides and shifts in topography.
These rapid changes in the landscape posed challenges for environmental monitoring and required advanced techniques to assess the new hydrological conditions and the state of surface waterbodies in the aftermath.

In an experimental analysis, we use IWD-QUEEN to compare inland waterbody changes in the Mekong River Basin before and after the seismic event. 
As shown in Figure \ref{fig: mekong earthquake}, the three-week period prior to the earthquake event (from 3/7 to 3/27) is taken as a reference, while the three-week period following the earthquake event (from 3/28 to 4/17) is used to observe the impact of the earthquake on inland waterbody changes.
No significant differences are observed on a large scale between the two periods. 
Therefore, we select three representative regions with different characteristics: Lake Lam Pao (103.5\textdegree E, 16.7\textdegree N), the Vang Naxay Wetland (103.2\textdegree E, 18.7\textdegree N), and the agricultural area of Ban Don (102.1\textdegree E, 18.6\textdegree N), to further compare the detailed changes before and after the earthquake. 

\begin{itemize}
    \item Black Box Region (103.25\textdegree E to 103.75\textdegree E, 16.5\textdegree N to 17\textdegree N):\\
    This region corresponds to Lake Lam Pao.
    Before the earthquake, the detection points are more numerous and densely concentrated in the upper portion of the lake. 
    However, after the earthquake, the detection of surface water extends into the lower portion of the lake. 
    Although the overall contour of Lake Lam Pao became more complete post-earthquake, the total surface water area detected decreased.

    \item Blue Box Region (103.25\textdegree E to 103.75\textdegree E, 18.5\textdegree N to 19\textdegree N):\\
    For the Vang Naxay Wetland, the distribution of detection points before and after the earthquake remained very similar. 
    Located between the Phu Bia Mountains, the wetland area is likely constrained by the surrounding topography, and no significant short-term changes are observed in this region.

    \item Green Box Region (102\textdegree E to 102.5\textdegree E, 18.5\textdegree N to 19\textdegree N):\\
    In the agricultural region of Ban Don, we observed considerable differences in the detection of agricultural contours before and after the earthquake.
    Before the earthquake, the agricultural area is primarily concentrated in the upper-left portion, along the tributaries of the Nam Lik River. 
    However, after the earthquake, the detected points do not show the northeast-southwest orientation typical of the agricultural land in this area.
\end{itemize}

\begin{table}[t]
\caption{Detection rates of the selected two time durations in the defined box areas.}
\centering
\begin{tabular}{l|c|c}
\hline\hline
& Before Earthquake & After Earthquake \\
\hline
Black Box & 42 / 349 = 12\% & 16 / 272 = 5.9\% \\
\hline
Blue Box  & 23 / 97 = 23.7\% & 17 / 75 = 22.7\% \\
\hline
Green Box & 18 / 119 = 15.1\%  & 7 / 82 = 8.5\%    \\
\hline
\hline
\end{tabular} \label{tab: mekong earthquake}
\end{table}

The quantitative analysis shown in Table \ref{tab: mekong earthquake} indicates that the detection rates for the black box (natural lakes) before and after the earthquake are 12\% and 5.9\%, respectively.
For the blue box (wetlands), the detection rates are 23.7\% and 22.7\%, and for the green box (agricultural land), the rates are 15.1\% and 8.5\%.
These results indicate that detection rates in areas with higher population development, particularly agricultural regions, exhibit more significant changes before and after the earthquake. 
It is hypothesized that, following the earthquake, deteriorating water quality led to increased turbidity, prompting the need to open the Dam Lam Pao sluice gates to prevent sedimentation. 
Consequently, the detection rate in this area decreased to only half of that observed prior to the earthquake. 
Additionally, agricultural areas may have experienced accelerated soil loosening due to seismic activity, leading to increased water loss and a reduction in surface water area after the earthquake. 
In contrast, the wetland areas located between the mountains, which remain undeveloped, exhibited greater environmental resilience due to effective water conservation, resulting in a more stable water presence during the seismic event.

\section{Conclusion and Future Work}\label{sec: conclusion}

In this paper, we have employed the strong classification ability of quantum deep learning to effectively detect inland waterbody regions. 
To make this feasible, we first encode the CYGNSS DDM data into compact features using a customized transformer, so that the transformer-encoded DDM can be represented using limited quantum bits in the near-term quantum computer. 
Besides the provable quantum full expressibility (cf. Theorem \ref{theorem: fully expressibility}) for ensuring the lightweight model, we also experimentally demonstrate the strong entanglement mechanism does help further upgrade the detection performance.
This novel technology, termed IWD-QUEEN, has achieved state-of-the-art IWD performance both quantitatively and qualitatively, with comprehensive case studies (e.g., waterbody retrieval among varying seasonal conditions and 2025 seismic event, etc.).
Furthermore, considering the potential users of traditional computers, and given the high efficacy of IWD-QUEEN, we also mimic the architecture of IWD-QUEEN to develop its non-quantum counterpart called IWD-Transformer. 
Though with some performance degradation, IWD-Transformer also yields superior detection results over existing baseline methods.

Therefore, this is the first paper that proposes the first practical and millisecond-level IWD solution, which requires only the easily accessible CYGNSS DDM data. 
Even with such a simple mechanism and data requirement, our IWD solution is able to effectively reveal the expansion of river networks in deep forest areas.
As the data facilitates effective estimation of physical parameters on the Earth's surface with high temporal resolution and wide spatial coverage, IWD-QUEEN can be used for near-real-time monitoring of waterbody dynamics in the future.
Technically, we also aim to explore other quantum-inspired remote sensing applications (e.g., satellite-based road/street detection, and soil moisture detection) as future research lines.

%\appendix

\renewcommand{\thesubsection}{\Alph{subsection}}
\bibliography{ref}

\begin{IEEEbiography}[{\resizebox{0.9in}{!}{\includegraphics[width=1in,height=1.25in,clip,keepaspectratio]{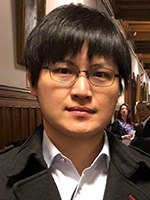}}}]
	{\bf Chia-Hsiang Lin}
	(S'10-M'18-SM'24)
received the B.S. degree in electrical engineering and the Ph.D. degree in communications engineering from National Tsing Hua University (NTHU), Taiwan, in 2010 and 2016, respectively.
From 2015 to 2016, he was a Visiting Student of Virginia Tech,
Arlington, VA, USA.

He is currently an Associate Professor with the Department of Electrical Engineering, and also with
the Miin Wu School of Computing,
National Cheng Kung University (NCKU), Taiwan.
Before joining NCKU, he held research positions with The Chinese University of Hong Kong, HK (2014 and 2017),
NTHU (2016-2017),
and the University of Lisbon (ULisboa), Lisbon, Portugal (2017-2018).
He was an Assistant Professor with the Center for Space and Remote Sensing Research, National Central University, Taiwan, in 2018, and a Visiting Professor with ULisboa, in 2019.
His research interests include network science,
quantum computing,
convex geometry and optimization, blind signal processing, and imaging science.

Dr. Lin received the Emerging Young Scholar Award (The 2030 Cross-Generation Program) from National Science and Technology Council (NSTC), from 2023 to 2027,
the Future Technology Award from NSTC, in 2022,
the Outstanding Youth Electrical Engineer Award from The Chinese Institute of Electrical Engineering (CIEE), in 2022,
the Best Young Professional Member Award from IEEE Tainan Section, in 2021,
the Prize Paper Award from IEEE Geoscience and Remote Sensing Society (GRS-S), in 2020,
the Top Performance Award from Social Media Prediction Challenge at ACM Multimedia, in 2020,
and The 3rd Place from AIM Real World Super-Resolution Challenge at IEEE International Conference on Computer Vision (ICCV), in 2019.
He received the Ministry of Science and Technology (MOST) Young Scholar Fellowship, together with the EINSTEIN Grant Award, from 2018 to 2023.
In 2016, he was a recipient of the Outstanding Doctoral Dissertation Award from the Chinese Image Processing and Pattern Recognition Society and the Best Doctoral Dissertation Award from the IEEE GRS-S.
\end{IEEEbiography}

\begin{IEEEbiography}[{\resizebox{1in}{!}{\includegraphics[width=1in,height=1.25in,clip,keepaspectratio]{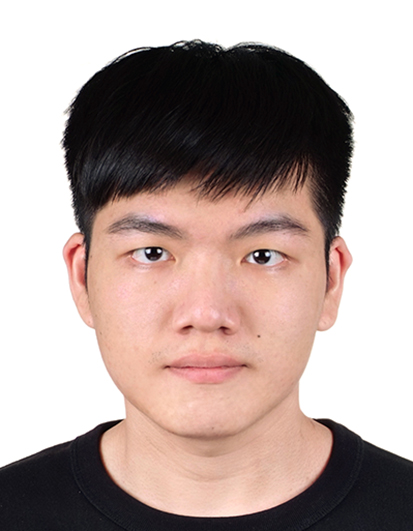}}}]
    	{\bf Jhao-Ting Lin}
		(S'20)
received his B.S. degree from the Department of Communications, Navigation and Control Engineering, National Taiwan Ocean University, Taiwan, in 2020.

He is currently a Ph.D. student affiliated with the Intelligent Hyperspectral Computing Laboratory, Institute of Computer and Communication Engineering, Department of Electrical Engineering, National Cheng Kung University, Taiwan. 
His research interests include convex optimization, deep learning, signal processing, quantum computing, and hyperspectral imaging.

He has received some highly competitive student awards, including the 2022 and 2024 Pan Wen Yuan Award from the Industrial Technology Research Institute (ITRI), Taiwan.
He has been selected as a recipient for the Ph.D. Students Study Abroad Program from the National Science and Technology Council (NSTC), Taiwan, for visiting the Okinawa Institute of Science and Technology in 2025.
\end{IEEEbiography}

\begin{IEEEbiography}[{\resizebox{1in}{!}{\includegraphics[width=1in,height=1.25in,clip,keepaspectratio]{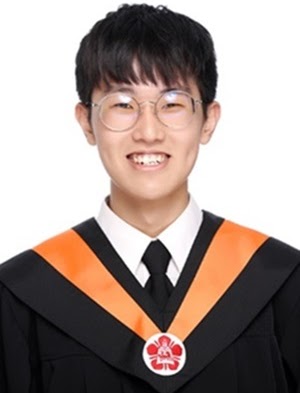}}}]
    	{\bf Po-Ying Chiu} (S'25)
received his B.S. degree from the Department of Systems and Naval Mechatronic Engineering, National Cheng Kung University, Taiwan, in 2024.

He is currently an M.S. student affiliated with the Intelligent Hyperspectral Computing Laboratory, Institute of Computer and Communication Engineering, Department of Electrical Engineering, National Cheng Kung University, Taiwan. 
His research interests include convex optimization, deep learning, signal processing, computer vision, quantum computing, and hyperspectral imaging.

\end{IEEEbiography}

\begin{IEEEbiography}[{\resizebox{1in}{!}{\includegraphics[width=1in,height=1.25in,clip,keepaspectratio]{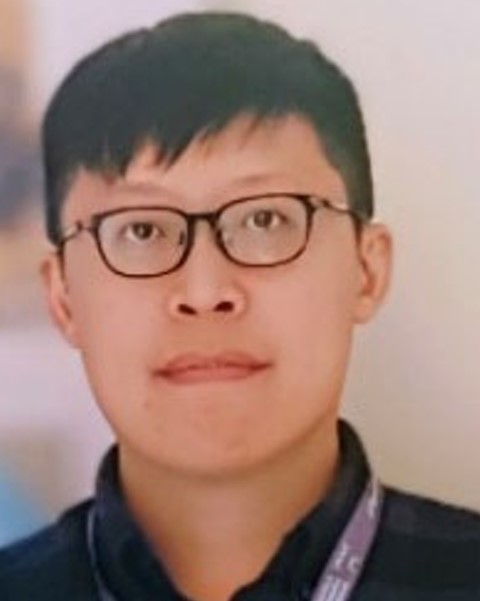}}}]
    	{\bf Shin-Ping Chen}
received the Ph.D. degree from the Institute of Space Science, National Central University (NCU), Taoyuan City, Taiwan, in 2017.

His thesis was a cooperative effort between NCU and the Department of Physics and Astronomy, George Mason University (GMU), Fairfax, VA, USA, on the modeling of global L-band scintillation observed by radio occultation satellites. 
He is currently an Assistant Research Fellow with the Department of Earth Sciences, National Cheng-Kung University (NCKU), Tainan City, Taiwan. 
His research interests include monitoring and modeling the influence of plasma irregularity to the HF communications and Global Navigation Satellite System (GNSS) applications.
\end{IEEEbiography}

\begin{IEEEbiography}[{\resizebox{1in}{!}{\includegraphics[width=1in,height=1.25in,clip,keepaspectratio]{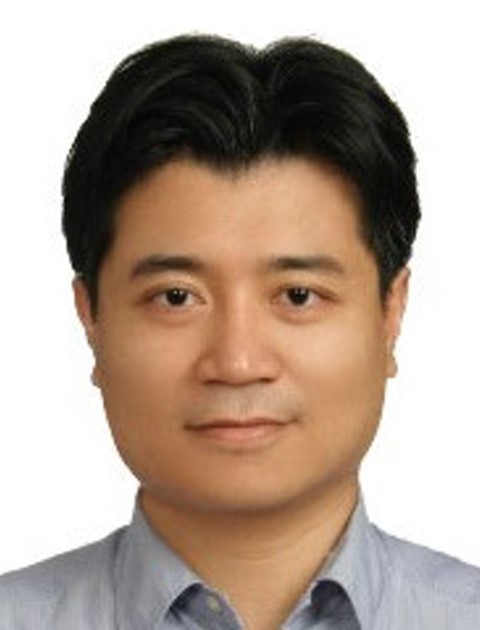}}}]
    	{\bf Charles C. H. Lin}
received the Ph.D. degree from the Institute of Space Science, National Central University (NCU), Taoyuan City, Taiwan, in 2005.

His thesis was a cooperative effort between NCU and the National Center for Atmospheric Research (NCAR), Boulder, CO, USA, on the theoretical modeling of thermosphere, ionosphere and plasmasphere modeling and magnetic storms. 
He is currently a Distinguished Professor with the Department of Earth Sciences, National Cheng-Kung University, Tainan City, Taiwan. 
His research interests include Global Navigation Satellite System (GNSS) applications in the upper atmosphere observations, ionospheric modeling, and data assimilation for global and regional scales.
\end{IEEEbiography}

\end{document}